\documentclass[11pt]{article}
\usepackage{amsmath}
\usepackage{amssymb}
\usepackage{amsfonts}
\usepackage{graphicx}
\usepackage{color}

\definecolor{WildStrawberry}{cmyk}{0, 0.96, 0.39, 0}
\definecolor{sky}{cmyk}{0.8, 0, 0.5, 0}

\numberwithin{equation}{section}
\begin{document}

\title{Geometry and stability of spinning branes in AdS gravity}
\author{Jos\'{e} D. Edelstein$\,^{a,b}$, Alan Garbarz$\,^{a,c}$, Olivera Miskovic$\,^{d}$ and Jorge Zanelli$\,^{b}$ \medskip\\
$^{a}${\small \emph{Department of Particle Physics and IGFAE, University of Santiago de Compostela,}}\\
{\small \emph{\ E-15782 Santiago de Compostela, Spain.}}\\
$^{b}${\small \emph{Centro de Estudios Cient\'\i ficos (CECS), Casilla 1469, Valdivia, Chile.}}\\
$^{c}${\small \emph{Departamento de F\'{\i}sica, Universidad de Buenos Aires, FCEN-UBA,}}\\
{\small \emph{\ Ciudad Universitaria, Pabell\'{o}n 1, 1428, Buenos Aires, Argentina.}}\\
$^{d}${\small \emph{Instituto de F\'{\i}sica, Pontificia Universidad Cat\'{o}lica de Valpara\'{\i}so,}}\\
{\small \emph{Casilla 4059, Valpara\'{\i}so, Chile.}}\\
{\small \texttt{jose.edelstein@usc.es, alan@df.uba.ar, olivera.miskovic@ucv.cl, z@cecs.cl}}}
\maketitle

\begin{abstract}
The geometry of spinning codimension-two branes in AdS spacetime is analyzed in three and higher dimensions. The construction of non-extremal solutions is based on identifications in the covering of AdS space by isometries that have fixed points. The discussion focuses on the cases where the parameters of spinning states can be related to the velocity of a boosted static codimension-two brane. The resulting configuration describes a single spinning brane, or a set of intersecting branes, each one produced by an independent identification. The nature of the singularity is also examined, establishing that the AdS curvature acquires one in the form of a Dirac delta distribution. The stability of the branes is studied in the framework of Chern-Simons AdS supergravity. A class of branes, characterized by one free parameter, are shown to be stable when the BPS conditions are satisfied. In 3D, these stable branes are extremal, while in higher dimensions, the BPS branes are not the extremal ones.
\end{abstract}

\newpage
\tableofcontents

%%%%%%%%%%%%%%%%%% [1]
\section{Introduction}
%%%%%%%%%%%%%%%%%%

The study of extended sources in gravitational theories has a rich literature behind it. A full understanding of the spectrum is
necessary, among other things, to attain a consistent partition function. A notable example of this assertion is the case of anti-de Sitter (AdS) gravity in three dimensions with negative cosmological constant, whose spectrum contains BTZ black holes \cite{BTZ} and conical defects (well-behaved naked singularities that admit the interpretation of point particles) \cite{DJtH}, but no propagating gravitons. Still, the Euclidean path integral with only smooth manifold contributions does not lead to a sensible quantum gravity partition function; seemingly, there may be additional contributions that require the introduction of point-like sources \cite{MW}. This is precisely the class of configurations we aim to explore in the present article.

Any configuration in three-dimensional AdS gravity that has constant curvature arises from identifications of AdS space by a
discrete one-parameter subgroup of $SO(2,2)$ generated by a Killing vector \cite{BHTZ}. The construction of black holes by
identifications in AdS space can be extended to higher dimensions. If the identifications do not have fixed points, they produce {\it topological} AdS black holes. They have been constructed, for instance, in $D=4$ \cite{ABHP}, $D=5$ \cite{BGM}, and higher
dimensions \cite{MaxB}. On the other hand, if the identifications do have fixed points, they produce cone-like singularities whose
consistent interpretation demands the introduction of explicit source terms in the Lagrangian, localized at their singular {\it
loci}. Even in a three-dimensional theory, these singularities correspond to topological defects at the positions of external current insertions that couple in a gauge-invariant way to three-dimensional AdS gravity.  The angular deficit at the singularities account for the negative part of the BTZ mass spectrum \cite{MZone}, filling the gap between the vacuum and the massless BTZ black hole.

It may seem surprising that those identifications leading to cone-like or Dirac delta function singularities are particularly
interesting. They describe codimension-two branes in gravity theories in any dimension and, in particular, 0-branes in three
dimensions. By construction, these are locally constant curvature geometries except at isolated codimension-two hypersurfaces  whose presence modify the topology of loops and closed surfaces in the ambient spacetime. In all cases, the codimension-two branes represent localized sources not covered by event horizons. Since these configurations are built from geometrical considerations that do not rely on dynamics, they might a priori arise in many different theories. Nonetheless, they can be naturally accommodated within a suitable framework where they become extended sources for gauge theories of gravity, such as Chern-Simons (CS) theories \cite{EZ,Mora:2000me,Mora:2000ts}.

The nature of the conserved charges, as well as the stability of these branes, do depend on the action, since the fluctuations of
geometry around the solutions depend on the full set of field equations. In order to study the stability of the spinning branes,
we consider CS AdS supergravity actions \cite{TZ1,TZ2}. As we will discuss in this article, the framework is possibly broader than
this, though the only explicit realizations we have considered here are CS theories of (super)gravity \cite{Z,EZtwo}. The advantage of these gravitation theories is that they are \emph{gauge} theories of gravity and that the interaction between the external sources ($2p$-branes) and the gauge connection can be incorporated in the action in a background-independent, metric-free and gauge invariant way \cite{MZtwo}. This coupling between branes and non-abelian gauge connections generalizes the minimal coupling that describes the interaction between a point charge (0-brane) and the $U(1)$ gauge
connection \cite{Zmalda}. Besides, these theories have many amenable features such as an off-shell formulation with the symmetries dictated by the M-algebra \cite{HTZ}.

To be more concrete, the kind of gravitational theories under consideration in the present paper are those that can be written in
terms of a gauge connection. We will use the first order formalism, in which the gravitational fields are the vielbein 1-form, $e^a$, and the spin-connection 1-form, $\omega^{ab}$ ($a,b=0,1,\ldots,D-1$). They can be written altogether as components of the AdS gauge connection
\begin{equation}
\mathbf{A}=\frac{1}{\ell }\,e^a \mathbf{J}_a + \frac{1}{2}\,\omega ^{ab}\mathbf{J}_{ab}\,,
\label{AdS_connection}
\end{equation}
with the AdS generators $\mathbf{J}_{AB}=-\mathbf{J}_{BA}$ ($A,B=0,1,\ldots ,D$), where $\mathbf{J}_{aD}\equiv \mathbf{J}_a$. The Riemann curvature 2-form, $R^{ab}=d\omega ^{ab}+\omega _{\ c}^{a}\wedge \omega ^{cb}$, and the torsion 2-form, $T^a = de^a +\omega _{\ b}^a \wedge e^b$, are the components of the AdS curvature (field-strength),
\begin{equation}
\mathbf{F}\equiv d\mathbf{A}+\mathbf{A}\wedge \mathbf{A}=\frac{1}{\ell }\,T^a\, \mathbf{J}_a + \frac{1}{2}\,\left( R^{ab}+\frac{1}{\ell ^{2}}\,e^{a}\wedge e^{b}\right)\, \mathbf{J}_{ab}\,.
\label{AdS_curvatrure}
\end{equation}
The flat connection, $\mathbf{F}=0$, corresponds to a torsion-free, global AdS$_D$ spacetime.

We are interested in configurations defined on a $D$-dimensional manifold $\mathbb{M}_D$  described by the equation
\begin{equation}
\mathbf{F}=\mathbf{j}\, ,
\label{F=j}
\end{equation}
where the sources $\mathbf{j}$ are localized at codimension-two surfaces embedded in the $D$-dimensional spacetime. The 2-form currents must be covariantly conserved, $D(\mathbf{A})\,\mathbf{j}=0$, and they represent ($D-3$)-branes sweeping codimension-two histories. Since $\mathbf{F}$ vanishes almost everywhere, except on some submanifold of measure zero, these branes produce locally AdS spacetimes, that could be solutions of any gravitation theory with negative cosmological constant that admits global AdS solutions. Apart from CS AdS gravity, these include standard Einstein-Hilbert AdS gravity, or higher-dimensional generalization of General Relativity described by Lovelock gravities \cite{Lovelock}.

A similar procedure was employed to obtain locally flat geometries by identifications of Minkowski space in the framework of Einstein-Hilbert gravity, leading to stationary, axisymmetric solutions with sources in the form of rods along a line \cite{Harmark} (for a review, see \cite{PP}).

As mentioned above, in some gravities it might be difficult to identify a coupling in the action whose field equations possess $\mathbf{F}=\mathbf{j}$ as a particular solution. In $D=2n+1$ dimensions, there is at least one gauge-invariant coupling that gives rise to this kind of interaction with a codimension-two source,
\begin{equation}
I_{\text{int}}=-\kappa \int\limits_{\mathbb{M}_{2n+1}}\left\langle C_{2n-1}(\mathbf{A})\wedge\mathbf{j}\right\rangle \,.  \label{interaction}
\end{equation}
Here, $\kappa $ is a coupling constant, $\left\langle \cdots \right\rangle $ is the trace taken in the algebra of some gauge group that contains AdS and $C_{2n-1}(\mathbf{A})$ is the Chern-Simons form that satisfies $d\left\langle C_{2n-1}(\mathbf{A})\wedge\mathbf{j}\right\rangle =\frac{1}{n}\left\langle \mathbf{F}^n \wedge \mathbf{j} \right\rangle $, where $\mathbf{F}^n$ denotes the wedge product of $n$ field-strength 2-forms. This coupling can be added, for example, to ($2n+1$)-dimensional CS AdS gravity,
\begin{equation}
I_{\text{CS}}=\kappa  \int\limits_{\mathbb{M}_{2n+1}}\left\langle C_{2n+1}(\mathbf{A})\right\rangle \,.
\end{equation}
The resulting field equations, $\mathbf{F}^{n-1}\wedge ( \mathbf{F}-\mathbf{j} ) =0$, have $\mathbf{F}=\mathbf{j}$ as a
particular solution.

For the sake of concreteness, whenever we need to specify an action (for example, in order to study the stability of the solution), we shall focus to CS AdS gravity $I[\mathbf{A},\mathbf{j}]=I_{\text{CS}}+I_{\text{int}}$, coupled to the sources in the form (\ref{interaction}).

%The paper is organized as follows. Section 2 deals with the 0-branes in three-dimensional theories. After reviewing the static case, a locally AdS spinning 0-brane is constructed by a number of methods that use different approaches. In all cases, the source that corresponds to the CS coupling has the form as the one in Refs.\cite{EZ,MZtwo}. It is also shown that Killing spinors are supported in BPS 0-brane configurations, and thereby they are stable. In section 3, the case of 2-branes in five-dimensional theories is discussed. All possible identifications in the embedding space are examined, finding that they lead to 2-brane solutions with one or two angular momenta. Several inequivalent cases that arise are studied in great detail. It is argued that spinning configurations with angular momentum associated to the internal azimuthal angle in the brane  are \emph{generically} intersections of several codimension-two branes, each one produced by one Killing vector field identification.  Finally, the existence of BPS configurations is explored. Section 4 generalizes these results to higher-dimensional codimension-two branes. Our conclusions and avenues for future research are presented in Section 5.

%%%%%%%%%%%%%%%%%%%%%%%%%%%%%%%%%%%%%%%%%%%%%%%% [2]
\section{Three-dimensional $0$-branes \label{ThreeD branes}}
%%%%%%%%%%%%%%%%%%%%%%%%%%%%%%%%%%%%%%%%%%%%%%%%

In this section we consider three-dimensional AdS space and show how a spinning $0$-brane can be obtained from identifications in this space. The construction is analogous to the one that yields the spinning black hole in three dimensions \cite{BHTZ}. This method was employed to construct the static 0-brane (point particle) in 3 dimensions \cite{MZone}, yielding particle solutions whose mass spectrum extends the black hole mass parameter to negative values.

The procedure of boosting the static $0$-brane, along the lines of \cite{MTZ}, provides a natural physical interpretation of the parameters that enter in the construction of the spinning branes.

%%%%%%%%%%%%%%%%%%%%%%%%%%%%%%%%%%%% [2.1]
\subsection{Static 0-brane in brief}
%%%%%%%%%%%%%%%%%%%%%%%%%%%%%%%%%%%%

Let us start by reviewing the construction of the static $0$-brane. AdS$_3$ space can be seen as a hyperboloid in $\mathbb{R}^{2,2}$, where the Cartesian coordinates $x^{A}=(x^{0},x^{1},x^{2},x^{3})$ are subjected to the constraint $x\cdot x\equiv \eta_{AB}\,x^{A}x^{B}=-\ell ^{2}$. Here, $\ell$ is the AdS radius and $\eta = \text{diag}(-1,1,1,-1)$. Using the parameterization
\begin{equation}
\begin{array}{ll}
x^{0}=A\cos \phi _{03}\,,\qquad & x^{1}=B\cos \phi _{12}\,, \\ [0.5em]
x^{3}=A\sin \phi _{03}\,, & x^{2}=B\sin \phi _{12}\,,
\end{array}
\label{3dstatic}
\end{equation}
the AdS$_{3}$ space corresponds to the surface
\begin{equation}
x\cdot x=B^{2}-A^{2}=-\ell ^{2}\,.
\label{AdS}
\end{equation}
The AdS$_{3}$ metric then reads
\begin{equation}
ds^{2}=\frac{\ell ^{2}\,dB^{2}}{B^{2}+\ell ^{2}}-\left( B^{2}+\ell^{2}\right) \,d\phi _{03}^{2}+B^{2}d\phi _{12}^{2}\,.
\label{AdS metric}
\end{equation}
Note that by unwrapping the $\phi_{03}$ coordinate, the global covering of AdS$_{3}$ in polar coordinates is obtained and closed time-like curves are eliminated. The static 0-brane can be constructed as a topological defect in the (1-2)-plane by introducing a deficit in the $\phi_{12}$ angle \cite{EGMZone,EGMZtwo}
\begin{equation}
\phi _{12}=a_{0}\,\phi _{0}\,,\quad \phi_{0}\simeq \phi _{0}+2\pi \,,\quad 0<a_{0}\leq 1\,.
\label{defect}
\end{equation}
The periodicity of $\phi_{12}$ is thus $2\pi a_{0}$. The metric, with this identification, reads
\begin{equation}
ds_0^2= \frac{dr^2}{\frac{r^2}{\ell ^2}+a_0^2}-\left( a_0^2 + \frac{r^2}{\ell ^2}\right) dt_0^2+r^2d\phi_0^2\,,
\label{static-polar}
\end{equation}
where $r=a_{0}B$ and $t_{0}=\ell\,\phi_{03}/a_{0}$. Compared with the BTZ black hole, the parameter $-a_{0}^{2}=M$ is seen to play the role of the (negative) mass. The spacetime geometry given by (\ref{static-polar}) has constant  Riemann curvature and vanishing torsion everywhere, except at $r=0$. On the other hand, the conical singularity at $r=0$ is the locus of the static $0$-brane, where infinite curvature is concentrated. This Dirac delta distribution has support in the center of the $r\phi$-plane, that is the (1-2)-plane in the covering space $\mathbb{R}^{2,2}$, and we shall call it the 2-manifold $\Sigma_{12}$. Including the singular point, the AdS curvature (\ref{AdS_curvatrure}) can then be written as $\mathbf{F}=\mathbf{j}$, where the source is given by the current 2-form
\begin{equation}
\mathbf{j} \equiv -2\pi a_0\,\delta (\Sigma _{12})\,\mathbf{J}_{12}\,.
\end{equation}
We have defined the Dirac delta distribution 2-form that is coordinate-independent,
\begin{eqnarray}
\delta (\Sigma _{12}) &=&\delta (x^1)\delta (x^2)\,dx^1 \wedge dx^2
=\frac{1}{2\pi }\,\delta (r)\,dr\wedge d\phi \,.
\end{eqnarray}
%
%%%%%%%%%%%%%%%%%%%%%%%%%%%%%%%%%%%%%%% [2.2]
\subsection{Spinning 0-brane \label{spinning 0}}
%%%%%%%%%%%%%%%%%%%%%%%%%%%%%%%%%%%%%%%

In the \textit{non-extremal} case, the massive spinning 0-brane can be obtained from the global AdS metric (\ref{AdS metric}) by an identification produced by a linear combination of the Killing vectors $\partial_{\phi_{03}}$ and $\partial_{\phi_{12}}$. This procedure closely follows the one described in Ref.\cite{BHTZ} to construct the BTZ black hole. Consider two angular deficits in the $x^0x^3$- and $x^1x^2$-planes in $\mathbb{R}^{2,2}$,
\begin{equation}
\phi _{03}\simeq \phi_{03}+2\pi b\,,\qquad \phi_{12}\simeq \phi_{12}+2\pi a\,.
\end{equation}
The choice of real constants $a$ and $b$ is not unique since these parameters can be shifted by integers. We choose $a,b \in (0,1]$ and note that $a=b=1$ means that there is no angular deficit. This identification in $\mathbb{R}^{2,2}$ corresponds to an identification in AdS$_3$ space by some angle $\phi$, what can be made explicit by redefining the coordinates as
\begin{equation}
\phi _{03}=b\phi +\frac{u_1 t}{\ell }\,,\qquad \phi _{12}=a\phi +\frac{u_2 t}{\ell}\,, \label{angles}
\end{equation}
where $\phi$ is periodic with period $2\pi$. The transformation is invertible if $a u_1-b u_2\neq 0$. Then, $\phi \simeq \phi+2\pi $ induces the following identification in $\mathbb{R}^{2,2}$,
\begin{equation}
x^{A}\simeq \left(
\begin{array}{cccc}
\cos 2\pi b & 0 & 0 & -\sin 2\pi b \\
0 & \cos 2\pi a & -\sin 2\pi a & 0 \\
0 & \sin 2\pi a & \cos 2\pi a & 0 \\
\sin 2\pi b & 0 & 0 & \cos 2\pi b%
\end{array}
\right) \,x^{A}\,,
\label{covering identification}
\end{equation}
or, in an infinitesimal form,
\begin{equation}
x^{A}\simeq x^{A}+\xi ^{A}\,,\qquad \xi ^{A}=\left( -2\pi b\,x^{3},-2\pi a\,x^{2},2\pi a\,x^{1},2\pi b\,x^{0}\right) \,.
\end{equation}
The identification is produced by the Killing vector
\begin{eqnarray}
\xi &=&\xi ^{A}\partial _{A}=2\pi a\,\mathbf{J}_{12}-2\pi b\,\mathbf{J}_{03}\,, \notag  \\ [0.5em]
&=&2\pi a\,\partial _{\phi _{12}}+2\pi b\,\partial _{\phi _{03}}\,,\label{nonextreml Killing}
\end{eqnarray}
where $\mathbf{J}_{AB}=x_{A}\partial _{B}-x_{B}\partial _{A}$ are generators of rotations in $\mathbb{R}^{2,2}$, and
\begin{equation}
\partial_{\phi_{12}}=\dfrac{\partial x^{A}}{\partial \phi_{12}}\,\partial_{A}=\mathbf{J}_{12}\,,\qquad \partial_{\phi_{03}} =\dfrac{\partial x^{A}}{\partial \phi_{03}}\,\partial_{A}=-\mathbf{J}_{03} \,.
\end{equation}
According to the classification given in Ref.\cite{BHTZ}, this Killing vector is of the type \textbf{I}$_c$ when $a\neq b$ ($a,b\neq 0$), and of type \textbf{II}$_b$ when $a=b\neq 0 $. Thus, this geometry belongs to a sector topologically different from the BTZ black hole, produced by identifications of type \textbf{I}$_b$ ($r_{+}\neq r_{-}$, $r_\pm \neq 0$) and type \textbf{II}$_a$ ($r_{+}=r_{-}\neq 0$).

Using the ortho-normality of the tangent vectors, $\partial _{A}\cdot \partial _{B}=\eta _{AB}$, the norm of the Killing vector $\xi$ reads
\begin{equation}
\left\Vert \xi \right\Vert ^{2}=(2\pi )^{2}\left[ \left( a^{2}-b^{2}\right)
B^{2}-b^{2}\ell ^{2}\right] \,,
\end{equation}
which is valid for any $a$ and $b$. However, notice that if $a^{2}\leq b^{2}$, the norm $\left\Vert \xi \right\Vert ^{2}$ is a time-like vector, and since it identifies different points in our geometry, it would lead to closed time-like curves. Therefore, we assume $a^{2}>b^{2}$,
\begin{equation}
\left\Vert \xi \right\Vert ^{2}=(2\pi )^{2}\left( a^{2}-b^{2}\right) \left( B^{2}-B_*^{2}\right) \,,\qquad B_*=\frac{b\ell }{\sqrt{a^{2}-b^{2}}} \,.
\label{3D_KV}
\end{equation}
After implementing the identification, the metric (\ref{AdS metric})
becomes
\begin{eqnarray}
ds^2 &=&\frac{\ell^2 dB^2}{B^2+\ell^2}+(a^2-b^2)(B^2-B_*^2)d\phi ^2 \nonumber \\
&&-\left( \frac{u_1^2-u_2^2}{\ell ^2}\,B^2+u_1^2\right) \, dt^2-2\left( \frac{au_2-bu_1}{\ell }\,B-u_1 b\ell \right)\, d\phi \,dt\,.
\label{boosted metric}
\end{eqnarray}
In sum, the metric (\ref{boosted metric}) is obtained from global AdS by the transformation (\ref{covering identification}) and
simultaneous identifications of the embedding angles $\phi_{03}$ and $\phi_{12}$. Alternatively, this metric  with $u_1=a$ and $u_2=b$ can be re-interpreted as the static $0$-brane with angular defect $a_0$, whose spin is introduced by boosting the azimuthal angle $\phi $ with the velocity $0\leq v<1$, similarly to the construction in Ref.\cite{MTZ} for the spinning charged
black hole.

Indeed, if $b^2 < a^2$, we can always write $a^2-b^2 \equiv a_0^2>0$, where $a_0 \in (0,1]$, and parameterize the constants $a$ and $b$ in terms of a hyperbolic angle $\eta$  as $a=a_0\cosh \eta $ and $b=a_0\sinh \eta $. Then, $\eta $ can be re-interpreted as the rapidity of some Lorentz transformation with velocity $v=\tanh \eta$, $0\leq v<1$, so that the original angular deficits are related to the boost velocity as
\begin{equation}
a=\frac{a_{0}}{\sqrt{1-v^{2}}}\,,\qquad b=\frac{va_{0}}{\sqrt{1-v^{2}}} \,.
\label{a0, omega}
\end{equation}
Now, the relations (\ref{angles}) with $u_1=a$ and $u_2=b$ are equivalent to a single Lorentz boost $( t_0,\phi _0) \rightarrow
(t,\phi )$ acting on the static $0$-brane,
\begin{equation}
t=\frac{t_{0}-\ell v\phi _{0}}{\sqrt{1-v^{2}}}\,,\qquad \phi =\frac{\phi_{0}-\frac{v}{\ell }\,t_{0}}{\sqrt{1-v^{2}}}\,.
\end{equation}
Note that this interpretation is possible only for $b^2 < a^2$ ($v\neq 1$) which means that the brane is not extremal, and
allows for the identification in the $(t, \phi)$-plane as $(t,\phi) \simeq (t, \phi+ 2\pi)$. The limit $v=0$ recovers the
static brane ($a=a_0$, $b=0$). Thus, $b$ is related to the angular momentum of the brane.

Although the original static metric describes a manifold with a conical singularity, the addition of angular momentum makes the
manifold regular, its curvature being constant everywhere \cite{MM}. This spacetime, however, has a \textit{causal
horizon}, the surface $B=B_*$, where the norm of the Killing vector field (\ref{3D_KV}) vanishes. The component  $g_{\phi \phi }=\left\Vert \xi \right\Vert ^{2}$ of the metric also vanishes there, though its positivity should guarantee that $\phi$ is an angle.

In the exterior region, $B>B_*$, the vector field $\xi$ is space-like and the causal structure is well-defined. In the interior region, $B<B_*$, $\xi$ becomes time-like allowing closed time-like curves. This region of spacetime can be removed by introducing a new radial coordinate $r$, such that $g_{\phi \phi }=r^2$ is always non-negative, or
\begin{equation}
B^{2}=\frac{r^2}{a_0^2}+B_*^2 \,,
\label{B}
\end{equation}
where the above transformation is valid for $B\geq B_*$. The removed region corresponds to $0<B< B_*$ and
$g_{\phi \phi}< 0$, where the boundary $g_{\phi \phi }=r^2 =0$ is shrunk to a point in the $r$-$\phi$ plane.  Note that the identification $B= B_*$ as a single point in the $r$-$\phi$ plane is not produced by the Killing vector. The region of interest, $B>B_*$, excludes the original singularity where the static $0$-brane lays. In spite of this, in what follows we shall show that there is a singular behavior at $r=0$ due to the identification of this circle (at fixed time) with a point. If this identification had not been performed, the resulting spacetime would have been regular, but with a geometry similar to the one of the spinning BTZ black hole at $r=0$ (see Appendix B of Ref.\cite{BHTZ}).

Once the region with closed time-like curves has been removed, the metric can be recast into the ADM form
\begin{equation}
ds^{2}=-N^{2}dt^{2}+\frac{dr^{2}}{N^{2}}+r^{2}\left( d\phi +N^{\phi }\,dt\right) ^{2} \,,
\label{Schwarzschild-like}
\end{equation}
where the lapse and shift functions are
\begin{equation}
N^{2}=a^{2}+b^{2}+\frac{r^{2}}{\ell ^{2}}+\frac{\ell ^{2}a^{2}b^{2}}{r^{2}} \,,\qquad N^{\phi }=-\frac{ab\ell }{r^{2}} \,.
\label{lapse-shift}
\end{equation}
Note that, with this parameterization, the metric is well-defined in the limit $a^2=b^2$, although the extremal case is obtained by a different identification in AdS space \cite{MZone}. However, this is a peculiarity of the three-dimensional metric only.

Even though the Lagrangian governing the dynamics of this $0$-brane has not been introduced, the existence of a nontrivial angular momentum can be established by calculating the angular velocity, $\Omega =-g_{\phi t}/g_{\phi \phi}$,  at the circle $r=Const$,
\begin{equation}
\Omega =-N^{\phi }=\frac{\ell ab\,}{r^{2}}\neq 0\,.
\end{equation}

The geometry of this asymptotically locally AdS spacetime is an analytic continuation of the $2+1$ black hole, where the parameters $a,b$ are continuations of the horizons $r_\pm$. By comparison with the $2+1$ black hole, the real parameters $a$ and $b$ can be related to the mass, $M$, and angular momentum, $J$, of the BTZ solution as
\begin{equation}
a\pm b = \sqrt{-M\pm \frac{J}{\ell}} \,,
\label{a,b-M,J}
\end{equation}
even without the knowledge of the Lagrangian. Thus, the metric (\ref{Schwarzschild-like}) describes a BTZ-like $0$-brane with a negative mass parameter, $M=-(a^{2}+b^{2})<0$,  boosted with respect to the static brane as $M=\frac{1+v^{2}}{1-v^{2}}\,M_{0}$. The parameter $b\neq 0$ is related to the angular momentum, $J=2ab\ell$.

%%%%%%%%%%%%%%%%%%%%%%%%%%%%%%%%%%%%%%%%%%%%%%%%% [2.3]
\subsection{Sources for a spinning $0$-brane \label{3D source}}
%%%%%%%%%%%%%%%%%%%%%%%%%%%%%%%%%%%%%%%%%%%%%%%%%

Let us summarize what we have done so far. An identification by a Killing vector of the pseudosphere $x\cdot x=-\ell ^2$  embedded in $\mathbb{R}^{2,2}$ amounts to a single identification by $\xi=2\pi(a\partial_{\phi_{12}}+b\partial_{\phi_{03}})$ in AdS$_{3}$.  The resulting manifold $\mathbb{M}'=$AdS$_{3}/\xi$ is described by the metric (\ref{boosted metric}). Since the identification is made by an isometry and is properly discontinuous, except at $B=0$, $\mathbb{M}'$ has constant curvature for $B>0$ and thus there is no curvature singularity. This manifold, however, contains closed time-like curves in the region  $B<B_*$ ($r<0$). Therefore, in order to have a causally well defined spacetime, the region $B<B_*$ must be removed by cutting along the surface  $B=B_*$, defined by $\left\Vert \xi \right\Vert =0$. Then all points that satisfy $B=B_*$ at fixed time are identified, producing a new manifold, $\mathbb{M}$, which has a naked singularity at $B=B_*$.

Another way of constructing the same geometry is the following. Take the AdS$_3$ spacetime (\ref{AdS metric}) and remove a portion of space $B<B_*$ (for some arbitrary $B_*$). Then identify the points of the resulting space with (\ref{covering identification}), where now $a$ and $b$ satisfy the relation with $B_*$ given by (\ref{3D_KV}). In this way, no closed time-like curves are produced since the region where they would appear was already cut out from space, and also there is no singularity. Now, the origin of the manifold is actually the circle $B=B_*$ ($r=0$) for fixed time, so by identifying all those points --which is \textit{not} done by a Killing vector-- a curvature singularity appears now at the origin. The question we want to analyze is what happens with the curvature at the point $r=0$ of $\mathbb{M}$ after this last identification.

The position of the source responsible for the singularity in spacetime is determined by the surface where the norm of the Killing vector vanishes. In the case at hand, we thus expect a source with a Dirac delta-like distribution of the form $\delta(|| \xi ||) \sim \delta (r)$. As shown below, this is indeed the case and there are no stronger singularities on the manifold, such as
$\delta (r)/r$, or $\partial_r\delta (r)$, etc.

In general, the singularity appears because the 1-form $d\phi$ is not exact on the whole manifold $\mathbb{M}$. Namely, at $r=0$, where $\phi$ is not defined, it is not true that $dd\phi =0$. Thus, we shall assume that $dd\phi =\Delta (r)\,dr\wedge d\phi $, where $\Delta (r)$ is some distribution that is zero when $r\neq 0$ and infinite when $r=0$. The static $0$-brane, for example, has $\Delta(r)=a_{0}\,\delta (r)$.

In order to identify the source and  nature of the singularity, one can construct the AdS connection using the vielbein and the
spin-connection for the metric (\ref{boosted metric}) in the region of interest $B>B_*$ ($r>0$),
\begin{eqnarray}
\mathbf{A} &=&\dfrac{\partial_r B \,dr}{\sqrt{B^{2}+\ell ^{2}}}\,\mathbf{J}_{13}+\frac{1}{\ell }\,\left[ B\left( b\mathbf{J}_{01}+a\mathbf{J}_{23}\right) +\sqrt{B^{2}+\ell ^{2}}\left( b\mathbf{J}_{03}-a\mathbf{J}_{12}\right) \right] d\phi   \nonumber \\ [0.5em]
&&+\frac{1}{\ell ^{2}}\,\left[ B\left( a\mathbf{J}_{01}+b\mathbf{J}_{23}\right) +\sqrt{B^{2}+\ell ^{2}}\left( a\mathbf{J}_{03}-b\mathbf{J}_{12}\right) \right] dt \,.
\label{A_AdS}
\end{eqnarray}
The curvature for $r>0$ is
\begin{equation}
\mathbf{F}=\left[ \frac{1}{\ell }\sqrt{\dfrac{r^{2}+\ell ^{2}a^{2}}{a^{2}-b^{2}}}\left( b\,\mathbf{J}_{03}-a\,\mathbf{J}_{12}\right) +\frac{1}{\ell }\sqrt{\dfrac{r^{2}+\ell ^{2}b^{2}}{a^{2}-b^{2}}}\left( a\,\mathbf{J}_{23}+b\,\mathbf{J}_{01}\right) \right] \Delta (r)\,dr\wedge d\phi \,.
\label{F}
\end{equation}
This form of $\mathbf{F}$ vanishes everywhere, with the possible exception at $r=0$. It has been recently claimed that spinning
branes require derivatives of the Dirac delta function (or $\delta (r)/r$) \cite{VVone,VVtwo}. Those arguments rely on the condition of vanishing torsion everywhere but, as discussed below, this requirement is not met by the present solution. In fact, as can be observed from (\ref{F}), the torsional parts (along $\mathbf{J}_{03}$ and  $\mathbf{J}_{23}$) have the same singular behavior as the curvature parts (along $\mathbf{J}_{01}$ and  $\mathbf{J}_{12}$). Thus, the configurations analyzed in Refs.\cite{VVone,VVtwo}, not obtained by identifications, correspond to different solutions compared to the ones considered here.

Let us assume a generic form for the singularity like $\Delta(r)=H(1/r)\delta (r)$, where $H$ is a power series in $1/r$ with
coefficients depending on $a$ and $b$. Then, requiring that the static $0$-brane is recovered in the limit $b\rightarrow 0$ implies that $H$ can be, at most, a linear combination of $\delta (r)$ and $\delta (r)/r$. Finally, expanding the expressions in (\ref{F}) as a series around $r=0$ leads to the source $\mathbf{j}=\mathbf{F}$ where
\begin{equation}
\mathbf{j}=\frac{1}{\sqrt{a^{2}-b^{2}}}\left[ ab\,\left(\mathbf{J}_{03}+\mathbf{J}_{23}\right) +b^{2}\,\mathbf{J}_{01}-a^{2}\,\mathbf{J}_{12}\right] \,\delta(r)\,dr\wedge d\phi \,. \label{J(c)}
\end{equation}
Clearly, this form of the source reproduces the static limit, but not the extremal one. The distribution $\Delta (r)$ can be calculated directly using the definition of the Riemann curvature, \textit{i.e.}, by parallel transport of a Lorentz vector $V^a$ along an infinitesimal contour around the point $r=0$. In this way, the Riemann curvature part of the source can be evaluated (see Appendix \ref{transport}), with the result
\begin{eqnarray}
\mathbf{j}_{\text{curvature}} = \frac{2\pi}{\sqrt{a^{2}-b^{2}}}\left(b^{2}\mathbf{J}_{01}-a^{2}\mathbf{J}_{12}\right)\delta (r)\,dr\wedge \frac{d\phi}{2\pi } \,. \label{j_R}
\end{eqnarray}
The torsional part of the source (along the generators $\mathbf{J}_{a3}$) cannot be obtained by the same parallel transport. However, comparing (\ref{j_R}) with (\ref{J(c)}), the $\mathbf{J}_{ab}$ terms match only if the distribution is identified as $\Delta(r)=\delta(r)$. The remaining components, along the generators $\mathbf{J}_{a3}$, correspond to the torsional part of
source,
\begin{eqnarray}
\mathbf{j}_{\text{torsion}} = \frac{2\pi ab}{\sqrt{a^2-b^2}}\,\left( \mathbf{J}_{03}+\mathbf{J}_{23}\right) \,\delta(r)\,dr\wedge \frac{d\phi}{2\pi }  \,.
\end{eqnarray}
It is then plain to see that the source carries no singularities stronger than $\delta(r)$.

The above method to calculate $\mathbf{j}$ is not easily generalized to higher dimensions because it calculates directly only
the Riemann curvature and the number of differential equations that need to be solved grows with the dimension.
The most precise way to determine the source is to use the fact that $\mathbf{F}$ is locally `pure gauge', so the AdS connection has the form
\begin{equation}
\mathbf{A}=g^{-1}dg \,,
\label{gdg}
\end{equation}
where $g$ is a group element of AdS. By solving this equation in $g$ with the AdS connection given by (\ref{A_AdS}), we obtain
\begin{equation}
g(t,B,\phi )=g_{0}\,e^{-\phi _{12}\mathbf{J}_{12}}e^{\phi _{03}\mathbf{J}_{03}}e^{p(B)\,\mathbf{J}_{13}} \,,
\label{g}
\end{equation}
where $\phi _{12}=a\phi +bt/\ell $, $\phi _{03}=b\phi +at/\ell $, we denote $p(B)=\sinh^{-1}(B/\ell)$ and $g_{0}$ is a constant element of the AdS group.

A non-trivial holonomy appears because $g$ is not single-valued. Indeed, the holonomy $g|_{\phi=2\pi}\,g^{-1}|_{\phi =0}$ is generated  by the Killing vector (\ref{nonextreml Killing}) as $e^{-\xi}$, up to a conjugation by a constant group element $g_{0}$. The result does not depend on the coordinates $B$ and $t$.

In order to calculate the curvature, one can look at the quantity $\oint_{\mathcal{C}_*}g^{-1}dg$, where $\mathcal{C}_*$ is a small circle of radius $B\simeq B_*$ around the causal horizon at constant $t$. We obtain
\begin{eqnarray}
\oint\limits_{\mathcal{C}_*}g^{-1}dg &=&-\frac{1}{\ell }\,\left( A_*\,\mathbf{J}_{12}-B_*\,\mathbf{J}_{23}\right) \oint\limits_{\mathcal{C}_*} \!d\phi _{12}\,+\frac{1}{\ell }\,\left( B_*\,\mathbf{J}_{01}+A_*\,\mathbf{J}_{03}\right) \oint\limits_{\mathcal{C}_*}\!d\phi _{03}\,+\oint\limits_{\mathcal{C}_*}\frac{dB}{A} \nonumber \\ [0.5em]
&=&-\frac{2\pi a}{\ell }\,\left( A_*\,\mathbf{J}_{12}-B_*\,\mathbf{J}_{23}\right) \,+\frac{2\pi b}{\ell }\,\left( B_*\,\mathbf{J}_{01}+A_*\,\mathbf{J}_{03}\right) \,,
\label{AU(1)}
\end{eqnarray}
where $A_*=\sqrt{B_*^{2}+\ell ^{2}}$. On the other hand, if $\Sigma _*$ is a surface whose boundary is $\mathcal{C}_*$, in a similar fashion it can be shown that $\int_{\Sigma _*}\mathbf{A}\wedge \mathbf{A}=0$, so that
\begin{equation}
\int\limits_{\Sigma _*}\!\mathbf{F}=\int\limits_{\Sigma _*}\!d\mathbf{A}=\oint\limits_{\mathcal{C}_*}g^{-1}dg \,.
\end{equation}
Note that a nontrivial result in the AdS curvature comes from $\mathbf{F}=g^{-1}ddg\sim dd\phi \neq 0$. Thus, due to the holonomy centered at $\left\Vert \xi \right\Vert =0$, or $B=B_*$ ($r=0$) in the $B\phi $-plane, we have
\begin{equation}
\int\limits_{\Sigma _*}\!\mathbf{F}=\frac{2\pi }{\sqrt{a^{2}-b^{2}}}\,\left( ab\mathbf{J}_{03}+ab\,\mathbf{J}_{23}+b^{2}\,\mathbf{J}_{01}-a^{2}\,\mathbf{J}_{12}\right) \,.
\end{equation}
Then, from (\ref{F}), $\Delta(r)=\delta(r)$, and from $\mathbf{j}=\mathbf{F}$ the source reads
\begin{equation}
\mathbf{j}=\frac{2\pi }{\sqrt{a^{2}-b^{2}}}\,\left[-a^{2}\mathbf{J}_{12}+b^{2}\mathbf{J}_{01}+ab\,\left( \mathbf{J}_{03} +\mathbf{J}_{23}\right)\right] \,\delta (r)\,dr\wedge \frac{d\phi}{2\pi } \,,  \label{j_holonomy}
\end{equation}
in perfect agreement with the result obtained before by other method. When $b=0$, this current produces the static brane,
$\mathbf{j}_{\text{static}}=-2\pi a\,\mathbf{J}_{12}\,\delta(r)\,dr\wedge \frac{d\phi}{2\pi } $. Also, when  $a=0$, the source simply vanishes. (Vanishing $a$ and $b$ are physically equivalent to $a=b=1$ in our choice of the range of these parameters.)

In the next subsection, we show that three-dimensional spinning $0$-branes can be stable.

%%%%%%%%%%%%%%%%%%%%%%%%%%%%%% [2.4]
\subsection{BPS spinning 0-branes}
%%%%%%%%%%%%%%%%%%%%%%%%%%%%%%

Locally AdS $0$-branes can be constructed in any gravity theory with negative cosmological constant where global AdS is an exact solution. In order to study their stability, one needs to know about the dynamics of the theory, and analyze fluctuations around the solution.  We thus need to provide a bulk Lagrangian, which we shall take to be CS AdS supergravity. This is a gauge theory whose supergroup is $OSp(p_{1}|2)$ $\times$  $OSp(p_{2}|2)$, and it contains $N=p_{1}+p_{2}$ supersymmetries. Apart from the vielbein and spin-connection, the super AdS connection contains additional bosonic components, that we call the bosonic CS matter.

In Ref.\cite{EGMZone}, it was shown that a static $0$-brane without bosonic CS matter is possibly unstable since it breaks all supersymmetries. Inclusion of $U(1)$ matter can stabilize the brane and turn it into a BPS state by preserving some supersymmetries in an extremal, charged, static case. Here we show that an extremal spinning $0$-brane can be a BPS state even without bosonic matter, as first found in Ref.\cite{IT} in the special case of $p_1=p_2=1$. In general, the number of preserved supersymmetries is determined by the number of Killing spinors $\epsilon_{I}^{\pm }$, each component `$+$' or `$-$ transforming as a vector in one copy of $OSp(p|2)$ labeled by the indices $I=1,\ldots, p$.

To find these spinors, we will make use of the static case analyzed in detail in Ref.\cite{EGMZone}, because both cases (spinning and static) have locally the same form. Thus, in the background of an uncharged  0-brane and with a suitable representation of the generators, the Killing spinor equation for each copy of $OSp(p|2)$ has the form
\begin{equation}
D_{\pm }(\mathbf{A})\epsilon _{I}^{\pm }=\left[ d-\left( \frac{1}{4}\epsilon_{\ bc}^{a}\,\omega ^{bc}\pm \frac{1}{2\ell}\, e^{a}\right)  \Gamma_{a} \right] \epsilon _{I}^{\pm }=0 \,,
\end{equation}
where $\Gamma_{a}$ are three-dimensional matrices satisfying the Clifford algebra (we consider only one of the two inequivalent representations of $\Gamma $-matrices, $c=1$ \cite{Coussaert:1993jp,MZone}). The vielbein and spin-connection are given in Eq.(\ref{A_AdS}). Then, a general solution for the Killing spinor is \cite{EGMZone}
\begin{equation}
\epsilon_I^{\pm}=e^{\pm \frac{1}{2}\,p(B) \Gamma_1} e^{\pm\frac{1}{2} \left(\phi_{12}-\phi_{03}\right) \Gamma_0}\chi_I^\pm  \,,
\end{equation}
where $\chi_I^\pm$ is a constant spinor fulfilling the chirality projection
\begin{equation}
\Gamma_0\,\chi_I^{\pm}=i\chi_I^{\pm} \,.
\label{projection}
\end{equation}
This Killing spinor is also \textit{globally} well-defined if it satisfies periodic or anti-periodic boundary conditions for $\phi \simeq \phi +2\pi $, \textit{i.e.}, $\phi_{12}\simeq \phi_{12}+2\pi a$ and $\phi_{03}\simeq \phi_{03}+2\pi b$. In consequence, $\epsilon_I^{\pm} \left(\phi +2\pi \right) =\pm \epsilon_I^{\pm} \left(\phi \right)$ implies the extremality condition
\begin{equation}
a-b=n\in \mathbb{Z} \,.
\end{equation}
When $b=0$ (static case), the only possibility to have this condition satisfied is for global AdS ($a=1$). For the spinning
0-brane, $a,b\in \left(0,1\right) $, the BPS configuration can exist even without additional bosonic matter, because the angular momentum plays the role of a $U(1)$ field. This is an accident of three dimensions only, where the metric admits the limit $a\rightarrow b$, even though the brane constructions for $a=b$ and $a\neq b$ differ \cite{MZone}.

These BPS states preserve $N/2$ supersymmetries; a half is projected out by the condition (\ref{projection}). The result can be generalized to include charged 0-branes, as well.

%%%%%%%%%%%%%%%%%%%%%%%%%%%%% [3]
\section{Five-dimensional 2-branes}
%%%%%%%%%%%%%%%%%%%%%%%%%%%%%

The construction of $2p$-branes by Killing vector identifications outlined above can be extended to higher dimensions by a procedure analogous to the static case \cite{EGMZone}. Here we present an explicit form of that construction in  $4+1$ dimensions.

The idea is to introduce naked singularities by making identifications with rotational Killing vectors in the four-dimensional spatial section of AdS in embedding space. The rotation generator $\mathbf{J}_{AB}$, that leaves invariant the $x^A$-$x^B$-plane, generates a conical singularity at the center, $x^A=x^B=0$.  Clearly, there are at most two independent identifications that can be performed simultaneously, generated by two commuting rotation generators in the four-dimensional spatial section, $\mathbf{J}_{AB}$ and $\mathbf{J}_{CD}$, with $(ABCD)$ a permutation of $(1234)$.

The resulting branes can also carry angular momentum, which would be the case if the identification is not restricted to be along the spatial section of AdS space, but is given by a generic element of the AdS$_5$ group  ($SO(4,2)$), corresponding to a boost on two planes simultaneously.  More precisely, the AdS$_5$ algebra has three commuting generators ($\mathbf{J}_{12}$,  $\mathbf{J}_{34}$ and $\mathbf{J}_{05}$, say) which allow for the introduction of three independent parameters in a locally AdS brane solution: the mass $M$, and two angular momenta $J_{\theta}$ and $J_{\phi}$.

%%%%%%%%%%%%%%%%%%%%%%% [3.1]
\subsection{General setting}
%%%%%%%%%%%%%%%%%%%%%%%

We will consider a codimension-two brane in a locally AdS spacetime, obtained by an identification with a Killing vector in global AdS$_{5}$. This brane is by definition the locus of points where the norm of the Killing vector field vanishes.

The AdS$_{5}$ spacetime can be viewed as the pseudo sphere $-(x^0)^2+ (x^1)^2 +(x^2)^2 +(x^3)^2 +(x^4)^2 -(x^5)^2= -\ell^2$ embedded in  $\mathbb{R}^{2,4}$, with metric $\eta_{AB}=$ diag$\left( -,+,+,+,+,-\right)$. A coordinate chart that describes this surface is given by
\begin{equation}
\begin{array}{lll}
x^{0}=A\cosh \tau \cos \phi_{05}\,,\quad  & x^{1}=B\cos \phi _{12}\,, \quad & x^{3}=A\sinh \tau \cos \phi _{34}\,, \\ [0.5em]
x^{5}=A\cosh \tau \sin \phi _{05}\,, & x^{2}=B\sin \phi _{12}\,, & x^{4}=A\sinh \tau \sin \phi _{34} \,,
\end{array}
\label{general embedding}
\end{equation}
where $A=\sqrt{B^{2}+\ell^2}$ and $B,\,\tau \in [ 0,\infty )$. These coordinates are chosen so that the $2$-brane is obtained by an identification in global AdS$_{5}$ along the azimuthal angles $\phi_{12}$, $\phi_{05}$ and $\phi_{34}$. Then $B$ becomes the radial coordinate outside the brane and $\tau$ is the radial coordinate in the worldvolume of the brane. The azimuthal angles $\phi_{12}$, $\phi_{05}$ and $\phi_{34}$ might have angular deficits in general, as the identifications are performed in the Euclidean planes $x^1$-$x^2$, $x^3$-$x^4$ and $x^0$-$x^5$, respectively.

In the coordinates (\ref{general embedding}), the metric takes the form
\begin{equation}
ds^{2}=\frac{\ell^2}{B^2+\ell^2} dB^2+B^2 d\phi_{12}^2 +\left(B^2 +\ell^2 \right) \left(d\tau^2 - \cosh^2\!\tau d\phi_{05}^2 +\sinh^2 \tau d\phi_{34}^2 \right).  \label{two spins}
\end{equation}
This metric corresponds to the covering of AdS$_5$ provided  $\phi_{05} \in \mathbb{R}$  (the unwrapped time coordinate) and $\phi _{12},\, \phi_{34}\in [0,2\pi]$ are periodic. Rescalings of these last two angles introduce angular deficits that characterize a particular identification. The most general identification, producing both linear defects and angular momenta in two planes, can be obtained by a general linear transformation between the embedding angles $\left(\phi_{12}, \phi_{05}, \phi_{34}\right)$ and some new coordinates in AdS$_5$. The new coordinates $\left( t, \phi, \theta \right)$, defined for $t\in \mathbb{R}$ and $\phi$, $\theta \in [0, 2\pi]$, are related to the original embedding coordinates by an invertible matrix $U$,
\begin{equation}
\left(
\begin{array}{c}
\phi _{05} \\
\phi _{12} \\
\phi _{34}
\end{array}
\right) =\left(
\begin{array}{ccc}
u_{1} & b_{1} & b_{2} \\
u_{2} & a_{1} & a_{2} \\
u_{3} & c_{1} & c_{2}
\end{array}
\right) \left(
\begin{array}{c}
t/\ell  \\
\phi  \\
\theta
\end{array}
\right) \\
= U \left(
\begin{array}{c}
t/\ell  \\
\phi  \\
\theta
\end{array}
\right)\,. \label{mix matrix}
\end{equation}
The identifications $(t/\ell,\phi,\theta) \simeq (t/\ell,\phi+2\pi,\theta)$ and $(t/\ell,\phi,\theta)\simeq (t/\ell,\phi,\theta +2\pi)$ are responsible for introducing angular defects. In the new coordinates, the metric (\ref{two spins}) can be cast in the familiar ADM form,
\begin{equation}
ds^2 =\frac{\ell^2 dB^2}{A^2}+ A^2 d\tau^2 -N^2 dt^2 +\gamma_{mn} \left(d\phi^m +N^m dt\right) \left( d\phi^n + N^n dt\right) \,,
\label{ADM angles}
\end{equation}
where $\phi^m =\left(\phi, \theta \right)$ are two independent azimuthal angles. The lapse $N$ and shift functions $N^m \equiv \left( N^{\phi },N^{\theta}\right)$, as well as the two-dimensional metric $\gamma_{mn}$, read
\begin{eqnarray}
N^2 & = & - \frac{1}{\ell^2}\, \left[ u_{2}^2 B^{2}+A^{2}\left( -u_{1}^2 \cosh ^{2}\tau +u_{3}^2 \sinh ^{2}\tau \right) \right] + \gamma_{mn}\, N^m N^n\,, \\ [0.5em]
N^m & = & \frac{1}{\ell}\, \gamma^{mn} \left[ u_{2}a_{n}B^{2}+A^{2}\left( -u_{1}b_{n}\cosh ^{2}\tau +u_{3}c_{n}\sinh ^{2}\tau \right) \right]\,, \\ [0.5em]
\gamma_{mn} & = & a_ma_nB^2 + A^2(-b_mb_n\cosh ^{2}\tau+c_mc_n\sinh ^{2}\tau)\,.
\end{eqnarray}
The off-diagonal components $g_{t\phi }$ and $g_{t\theta }$ are related to two possible nontrivial angular velocities of this spacetime. The components  $g_{tt}$, $g_{\phi \phi }$ and $g_{\theta \theta}$ are functions of the spacetime coordinates that might change signs and lead to a non-causal global structure, with regions containing closed time-like curves. In that case, those non-causal regions must be removed from the manifold, potentially generating singularities analogous to the central singularity in the $2+1$ black hole. $\Omega_m=-\gamma_{mn} N^n$ are angular velocities of the points at $B,\tau =Const$. Thus, the shift functions $N^n$ contain information about the non-vanishing angular momenta of these stationary, locally AdS brane geometries.

In this general setting, the action of the system has not been specified. Thus, the mass and angular momenta cannot be determined, as they are conserved charges defined by the Lagrangian via Noether's theorem. Furthermore, in order to obtain finite charges, boundary terms (and counterterms) must be added to the bulk action, that has been done in five-dimensional CS AdS gravity without torsion in \cite{Banados:2005rz}, with torsion in \cite{Banados:2006fe} and in higher-dimensional Lovelock AdS gravities in Ref.\cite{Kofinas:2007ns}. For some branes one can compare the asymptotic behavior of the metric with that of solutions in asymptotically AdS spaces with known mass and angular momenta in some standard theory. If the metrics match, the mass and angular momenta of the brane may be identified with those of the known solution.

On the other hand, unlike the angular momenta, the angular velocities $\Omega_m$ of the brane are kinematically determined by the geometry itself.

%%%%%%%%%%%%%%%%%%%%%%%%%%% [3.1.1]
\subsubsection{Identifications}
%%%%%%%%%%%%%%%%%%%%%%%%%%%

The periodicity of the coordinate $\phi$ in AdS space implies certain identifications of points in the embedding space,
\begin{equation}\label{phiidentification}
\begin{array}{cc}
\phi \simeq \phi +2\pi \quad \Leftrightarrow \quad  & \phi _{05}\simeq \phi_{05}+2\pi b_{1}\,, \\ [0.3em]
& \phi _{12}\simeq \phi _{12}+2\pi a_{1}\,, \\ [0.3em]
& \phi _{34}\simeq \phi _{34}+2\pi c_{1}\,,
\end{array}
\end{equation}
and similarly for $\theta$,
\begin{equation}\label{thetaidentification}
\begin{array}{cc}
\theta \simeq \theta +2\pi \quad \Leftrightarrow \quad  & \phi _{05}\simeq \phi _{05}+2\pi b_{2}\,, \\ [0.3em]
& \phi _{12}\simeq \phi _{12}+2\pi a_{2}\,, \\ [0.3em]
& \phi _{34}\simeq \phi _{34}+2\pi c_{2}\, .
\end{array}
\end{equation}
The ambiguity of the parameters $a_i,b_i,c_i$ under the addition of  integers can be eliminated restricting them to the range
$0<a_i,b_i,c_i\leq 1$, where the equality corresponds to the case with no angular deficit.  Using Eq. (\ref{general embedding}), these identifications, expressed infinitesimally as $x^A\simeq x^A+\xi_i^A$, $i=1,2$ in the embedding space, take the form
\begin{equation}
x^{A}\simeq \left(
\begin{array}{cccccc}
\cos 2\pi b_{i} & 0 & 0 & 0 & 0 & -\sin 2\pi b_{i} \\ [0.1em]
0 & \cos 2\pi a_{i} & -\sin 2\pi a_{i} & 0 & 0 & 0 \\ [0.1em]
0 & \sin 2\pi a_{i} & \cos 2\pi a_{i} & 0 & 0 & 0 \\ [0.1em]
0 & 0 & 0 & \cos 2\pi c_{i} & -\sin 2\pi c_{i} & 0 \\ [0.1em]
0 & 0 & 0 & \sin 2\pi c_{i} & \cos 2\pi c_{i} & 0 \\ [0.1em]
\sin 2\pi b_{i} & 0 & 0 & 0 & 0 & \cos 2\pi b_{i}
\end{array}
\right) x^{A}\,.
\end{equation}
They are generated by two linearly independent, commuting (in the sense of Lie brackets) Killing vector fields
\begin{equation}
\xi_{i} = 2\pi a_{i}\,\mathbf{J}_{12}-2\pi b_{i}\,\mathbf{J}_{05}+2\pi c_{i}\,\mathbf{J}_{34}\,, \label{5D_KV}
\end{equation}
whose norms are
\begin{equation}
\left\Vert \xi_i\right\Vert^2 = 4\pi^2\left(a_i^2 B^2 -b_i^2 A^2\cosh^2\!\tau + c_i^2A^2\sinh^2 \tau \right) \,.
\end{equation}
The Killing vectors (\ref{5D_KV}) act in the same spacetime, but in different spatial planes. Denoting them by $\xi_i=W_{(i)}^{AB}\mathbf{J}_{AB}$, the matrices $W_{(i)AB}$ are antisymmetric, block-diagonal, and have eigenvalues $\pm ia_i$, $\pm ib_i$ and $\pm ic_i$. The knowledge of three eigenvalues for each matrix is equivalent to knowing three Casimir invariants: quadratic ($W^{AB}W_{AB}$), cubic ($\epsilon ^{ABCDEF}W_{AB}W_{CD}W_{EF}$) and the quartic one ($W_{\ B}^{A}W_{\ C}^{B}W_{\ D}^{C}W_{\ A}^{D}$) or, in the case at hand, $a_i^2+b_i^2+c_i^2$, $a_i b_i c_i$ and $a_i^4+b_i^4+c_i^4$, respectively. The set of parameters is equivalent to the set of Casimir invariants, where only three are independent. These parameters are, therefore, gauge-invariant quantities related to the conserved charges $M$,  $J_\phi$ and $J_\theta$, where the explicit relations depend on the gravity action.

In order to provide a geometric interpretation for this general framework, we consider codimension-two branes that possess a static limit. These spinning branes can be obtained from the static ones by a Lorentz boost in a plane formed by the time coordinate and an azymuthal angle \cite{MTZ}. In five dimensions, there are two such angles, here called $\phi$ and $\theta$, so it is possible to do two idependent boosts, introducing two independent angular velocities.

Let us denote the coordinates of a static $2$-brane by $(t_0, B, \phi_0, \tau , \theta_0)$. In Ref.\cite{EGMZone}, it was shown that a static 2-brane can be produced by an angular deficit $2\pi (1-a_0 )$ in the $x^1$-$x^2$ plane through a rescaling of the angle $\phi$ and the time coordinate. In this case, the embedding angles read
\begin{equation}
\left(
\begin{array}{c}
\phi_{05} \\
\phi_{12} \\
\phi _{34}
\end{array}
\right) = U_0 \left(
\begin{array}{c}
t_0/\ell  \\
\phi _0 \\
\theta_0
\end{array}
\right),
\label{static angles}
\end{equation}
where $U_0 =$\;diag$(a_0, a_0, 1)$. The static metric has the form
\begin{equation}
ds_0^2= \frac{\ell^2 dB^2}{B^2+\ell^2}+B^2a_0^2\,d\phi_0^2+A^2\left( d\tau^2-\frac{a_0^2}{\ell^2}\cosh^2\!\tau
\, dt_0^2 +\sinh^2\!\tau \,d\theta_0^2\right) \,.  \label{5D static}
\end{equation}
The resulting geometry has a conical (Dirac delta) curvature singularity in the $B\phi_0$-plane \cite{EGMZone}. The coefficient $a_0$ is related to the mass of the brane, although the precise relation depends on the particular gravitational action that is assumed.

%%%%%%%%%%%%%%%%%%%%%%%%%%%%%%%%%%%%%% [3.1.2]
\subsubsection{AdS connection and group element}
%%%%%%%%%%%%%%%%%%%%%%%%%%%%%%%%%%%%%%%

The described 2-branes are locally AdS, so that the curvature vanishes locally, $\mathbf{F}=0$, which means constant negative Riemann curvature and vanishing torsion (\textit{cf.}, Eq. (\ref{AdS_curvatrure})). Thus, the AdS connection is ``locally pure gauge" and can be expressed in terms of an element of the AdS group $g$ as $\mathbf{A}=g^{-1}dg$. In order to calculate $g$ for the metric (\ref{two spins}), the vielbein can be chosen as
\begin{equation}
\begin{array}{ll}
e^{0}=A\,\cosh \tau \,d\phi _{05}\,,\medskip \qquad  & e^{3}=A\,d\tau \,, \\
e^{1}=\dfrac{\ell }{A}\,dB\,,\medskip  & e^{4}=A\,\sinh \tau \,d\phi _{34}\,, \\
e^{2}=B\,d\phi _{12}\,. &
\end{array}
\end{equation}
Since the torsion vanishes, the spin-connection can be algebraically obtained, and the AdS connection (\ref{AdS_connection}) has the form
\begin{eqnarray}
\mathbf{A} &=&g^{-1}dg=\frac{1}{A} dB\, \mathbf{J}_{15}+\frac{1}{\ell} \left( A\,\mathbf{J}_{35}-B\,\mathbf{J}_{13}\right) d\tau +\frac{1}{\ell}\left( B\,\mathbf{J}_{25}-A\,\mathbf{J}_{12}\right) d\phi_{12}  \notag \\ [0.5em]
&&+\frac{1}{\ell}\left( A\cosh \tau\, \mathbf{J}_{05}+B\cosh \tau\, \mathbf{J}_{01} +
\ell \sinh \tau\, \mathbf{J}_{03}\right) d\phi _{05}  \notag \\ [0.5em]
&&+\frac{1}{\ell}\left(A\sinh \tau\, \mathbf{J}_{45}-B\sinh \tau\, \mathbf{J}_{14}-\ell \cosh \tau\, \mathbf{J}_{34}\right) d\phi _{34}\,.
\label{AdS connection}
\end{eqnarray}
The group element of AdS$_{5}$ is
\begin{equation}
g=g_{0}\,e^{-\phi_{34}\mathbf{J}_{34}}e^{\phi_{05}\mathbf{J}_{05}}\,e^{-\phi_{12}\mathbf{J}_{12}}\,e^{\tau \mathbf{J}_{35}} e^{p(B) \mathbf{J}_{15}}\,,  \label{g general}
\end{equation}
where $p(B)=\ln \frac{B+A}{\ell}$ and $A=\sqrt{B^{2}+\ell^{2}}$, so that $A=\ell \cosh p$ and $B=\ell \sinh p$.

As in the three-dimensional case, the group element $g$ is not globally defined since $g|_{\phi^m=2\pi} \neq g|_{\phi^m=0}$, which is precisely what gives rise the brane geometry. With the explicit form of $g$, we can calculate the corresponding sources.

So far we have assumed a generic form for the identification matrix $U$. In the next subsections we study particular cases corresponding to interesting brane configurations.

%%%%%%%%%%%%%%%%%%%%%%%% [3.2]
\subsection{Spinning 2-brane}
%%%%%%%%%%%%%%%%%%%%%%%%

\begin{figure}[h]
\centering
\includegraphics[width=0.52\textwidth]{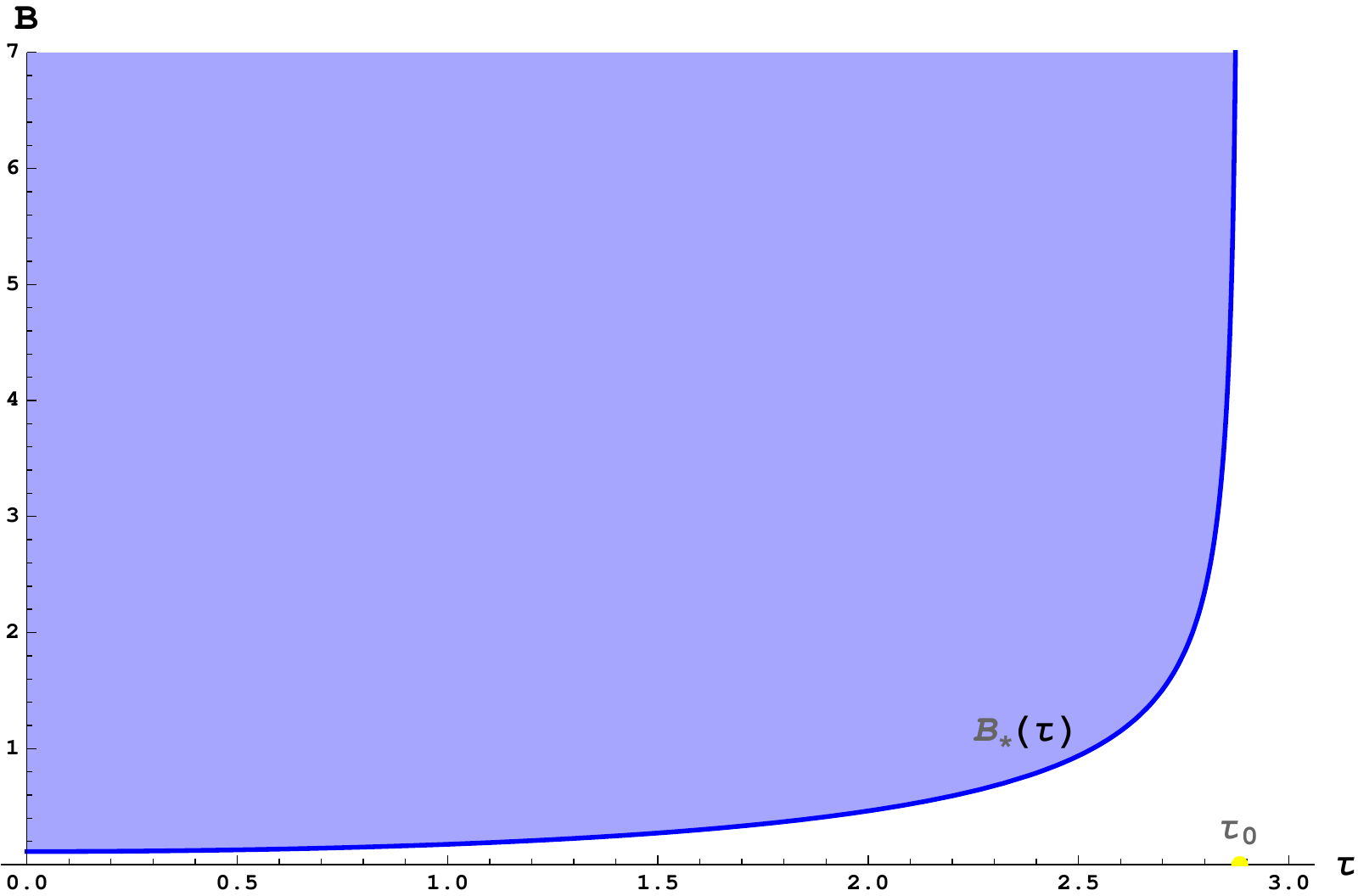}
\caption{The causal horizon $B_*(\tau)$. The physical region is the shaded area $B\geq B_*(\tau )$.}
\label{5Dbrane-phi}
\end{figure}

As discussed in the three-dimensional case, the spinning solution can be obtained by ``boosting" the static geometry, and this can also be done in five dimensions. We will first boost the static 2-brane (\ref{5D static}) in the (transverse) $\phi$ direction, by performing a Lorentz transformation $\Lambda_{\phi} (w)$ on the $t\phi$-plane, with velocity $0\leq w< 1$,
\begin{equation}
\Lambda_{\phi}(w): \qquad t=\frac{t_0-\ell w\phi_0}{\sqrt{1-w^2}}\,,\qquad \phi =\frac{\phi_0-\frac{w}{\ell} t_0}{\sqrt{1-w^2}}\,,\qquad \theta=\theta_0\,,
\end{equation}
represented by the matrix
\begin{equation}
\Lambda_{\phi}(w)=e^{-\zeta\mathbf{L}_{02}}=\left(
\begin{array}{ccc}
 \cosh\zeta & -\sinh \zeta & 0 \\
-\sinh\zeta & \cosh \zeta & 0 \\
0 & 0 & 1
\end{array}
\right) \,,  \label{Lambda1}
\end{equation}
where $w=\tanh \zeta$. The extreme case ($w=1$) is not included in this analysis. The angles in the embedding space are related to the AdS coordinates via the matrix $U=U_0\Lambda^{-1}_\phi$, so that
\begin{equation}
\left(
\begin{array}{c}
\phi_{05} \\
\phi_{12} \\
\phi_{34}
\end{array}
\right) =\left(
\begin{array}{ccc}
a_1 & b_1 & 0 \\
b_1 & a_1 & 0 \\
0 & 0 & 1
\end{array}
\right) \left(
\begin{array}{c}
t/\ell \\
\phi \\
\theta
\end{array}
\right) \,,  \label{J1 mixing}
\end{equation}
where the constants
\begin{equation}
a_1 = a_0\cosh \zeta\,, \qquad b_1 = a_0\sinh \zeta\,,
\end{equation}
satisfy
\begin{equation}
\det U=a_1^2-b_1^2=a_0^2>0\,.
\end{equation}
From $\phi \simeq \phi +2\pi $, the above boost can be seen to be equivalent to the identification
\begin{equation}
\phi_{12}\simeq \phi_{12}+2\pi a_1\,,\qquad \phi_{05}\simeq \phi_{05}+2\pi b_1\,,
\end{equation}
produced by the Killing vector
\begin{equation}
\xi_1=2\pi a_1\,\mathbf{J}_{12}-2\pi b_1\,\mathbf{J}_{05}\,.
\label{xi1}
\end{equation}
The norm of this vector is
\begin{equation}
\left\Vert \xi_{1}\right\Vert^{2}=4\pi^2\left(a_1^2 B^2-b_1^2 A^2\cosh^2\!\tau \right) \,,
\end{equation}
and it is non-negative only if
\begin{equation}
B\geq B_{*}(\tau )\,,\qquad \tau < \tau_0 \equiv \cosh ^{-1}\frac{a_{1}}{b_{1}}\,,
\label{xi_1>0}
\end{equation}
where
\begin{equation}
B_{*}(\tau )=\frac{b_1\ell \cosh\tau }{\sqrt{a_1^2-b_1^2\cosh^2\!\tau }}\,\,.  \label{B_h}
\end{equation}
The physical region corresponds to that where the norm of $\xi_1$ is positive, which requires $B>B_{*}$. Note that $B\geq B_{*}(0)=\ell b_1/a_0>0$ and hence this region does not contain the point $B=0$. The domain of the coordinates $B$ and $\tau$, $B\geq B_{*}(\tau )$, for which the spacetime is physical, is shown in Figure \ref{5Dbrane-phi}.
The horizon $B_{*}(\tau)$ can be identified with the origin of the radial direction, a point that shall be called $r=0$, by introducing a new radial coordinate $r=B-B_{*}(\tau)\geq 0$.

%%%%%%%%%%%%%
%%%%%%%%%%%%%

The explicit form of the metric for this solution in the ADM form (\ref{ADM angles}) is
\begin{equation}
ds^2=\frac{\ell ^2 dB^2}{B^2+\ell ^2}-N^2 dt^2+f^2\left( d\phi+N^\phi dt\right)^2+A^2\left( d\tau^2+\sinh^2\!\tau d\theta^2\right) \,,
\end{equation}
where the lapse and shift functions are
\begin{equation}
N^2=\frac{a_0^4 A^2 B^2 \cosh^2\!\tau}{\ell^2\left( a_1^2 B^2-b_1^2 A^2\cosh^2\!\tau \right) }\,,
\qquad N^\phi =\frac{a_1 b_1\left( B^2-A^2\cosh^2\!\tau \right) }{\ell \left( a_1^2 B^2-b_1^2 A^2\cosh ^2\tau \right) }\,,
\end{equation}
and we denoted $f^2=a_1^2 B^2-b_1^2 A^2\cosh^2\!\tau =\left\Vert \xi _1\right\Vert^2/4\pi^2$.

%%%%%%%%%%%%%%%%%%%%%%%%%%%%%%%%%%%%%%%% [3.2.1]
\subsubsection{Sources for a spinning 2-brane}
%%%%%%%%%%%%%%%%%%%%%%%%%%%%%%%%%%%%%%%%

The sources can be identified from the flux they produce. Since the curvature is given by the current (cf. Eq (\ref{F=j})), the flux can be computed integrating the connection around the singular points. Thus, starting from the AdS group element\ (\ref{g general}) for the angles (\ref{J1 mixing}), the source is obtained from $\oint g^{-1}dg$. Here, the only non-vanishing contour integrals are
\begin{equation}
\oint d\phi _{12}=2\pi a_{1}\,,\quad \oint d\phi _{05}=2\pi b_{1}\,,
\end{equation}
on infinitesimal loops around the Killing vector horizon $\left\Vert \xi_1\right\Vert =0$, given by $B=B_{*}(\tau)$,
\begin{eqnarray}
\int \mathbf{j} &=&\oint g^{-1}dg=\frac{2\pi a_1}{\ell}\left( B_{*}\, \mathbf{J}_{25}-\sqrt{B_{*}^2+\ell^2}\,\mathbf{J}_{12}\right)  \notag \\
&&+\frac{2\pi b_1}{\ell}\left(\sqrt{B_{*}^2+\ell^2}\cosh\tau\, \mathbf{J}_{05}+B_{*}\cosh\tau \,\mathbf{J}_{01}+\ell \sinh\tau \,\mathbf{J}_{03}\right) \,.
\end{eqnarray}%
Thus, since the angles $\phi_{12}$ and $\phi_{05}$ define $x^1$-$x^2$ and $x^0$-$x^5$ planes in the embedding space, respectively, the source can be written as
\begin{eqnarray}
\mathbf{j} &=&\frac{2\pi a_1}{\ell }\left( B_{*}\,\mathbf{J}_{25}-\sqrt{B_{*}^2+\ell^2}\, \mathbf{J}_{12}\right) \delta (\Sigma_{12})  \notag \\
&&+\frac{2\pi b_1}{\ell}\left(\sqrt{B_{*}^2+\ell^2}\cosh\tau \,\mathbf{J}_{05}+B_{*}\cosh\tau \,\mathbf{J}_{01}+\ell \sinh\tau \,\mathbf{J}_{03}\right) \delta (\Sigma_{05})\,,
\end{eqnarray}
where the Dirac delta 2-forms are
\begin{eqnarray}
\delta (\Sigma_{12}) &=&\delta (x^1)\delta(x^2) dx^1 \wedge dx^2\,,  \notag \\
\delta (\Sigma_{05}) &=&\delta (x^0)\delta(x^5) dx^0 \wedge dx^5\,,
\end{eqnarray}
and should be thought of as projected in AdS$_5$. Therefore, both deltas are proportional to $\delta(r) dr\wedge d\phi$, up to the corresponding Jacobians. The source describes one codimension-two brane generated by one Killing vector identification. The angular momentum is associated to the azimuthal angle $\phi $ outside the brane.

Next, we construct a brane with angular momentum associated to the interior azimuthal angle $\theta$, obtained by boosting this interior angle in the static brane.

%%%%%%%%%%%%%%%%%%%%%%%%%%%%% [3.3]
\subsection{Intersection of 2-branes}
\label{Intersection}
%%%%%%%%%%%%%%%%%%%%%%%%%%%%%

The static 2-brane (\ref{5D static}) can be boosted in the $\theta$ direction by a Lorentz transformation $\Lambda_{\theta} (v)$ in the $t_0$-$\theta_0$ plane, with velocity $0\leq v <1$,
\begin{equation}
\Lambda_{\theta}(v):\qquad t=\frac{t_{0}-\ell v \theta_0}{\sqrt{1-v^2}}\,,\qquad \phi =\phi_0\,,\qquad \theta =\frac{\theta_0 -\frac{v}{\ell} t_0}{\sqrt{1-v^2}}\,. \label{boost-v}
\end{equation}
Restricted to the space of the coordinates $(t_0/\ell, \phi_0, \theta_0)$, this transformation has the form
\begin{equation}
\Lambda_{\theta}=e^{-\eta \mathbf{L}_{04}}=\left(
\begin{array}{ccc}
\cosh \eta & 0 & -\sinh \eta \\
0 & 1 & 0 \\
-\sinh \eta & 0 & \cosh \eta
\end{array}
\right) \,,\quad \mathbf{L}_{04}=\left(
\begin{array}{ccc}
0 & 0 & 1 \\
0 & 0 & 0 \\
1 & 0 & 0
\end{array}
\right) \,,  \label{Lambda2}
\end{equation}
where the velocity is related to the hyperbolic angle, or rapidity, as $v=\tanh \eta$. The spinning geometry can be obtained by applying a Lorentz transformation to the corresponding coordinates of the static brane, $\Lambda_{\theta}: \left( t_0/\ell,\phi_0,\theta_0\right)  \mapsto \left( t/\ell, \phi, \theta \right)$.

The boosted geometry is the same as  (\ref{two spins}) after (\ref{phiidentification}) and (\ref{thetaidentification}), obtained
by taking the quotient of the five-dimensional pseudosphere by a Killing vector. The relation between the two results can be made explicit by combining the two steps, (\ref{static angles}) and (\ref{boost-v}), into a single transformation that expresses the embedding angles in terms of the coordinates of the spinning brane, where $U$ in (\ref{mix matrix}) is calculated from
\begin{equation}
\left(
\begin{array}{c}
\phi_{05} \\
\phi_{12} \\
\phi _{34}
\end{array}
\right) = U_0\,\Lambda_{\theta}^{-1} \left(
\begin{array}{c}
t/\ell  \\
\phi  \\
\theta
\end{array}
\right) = \left(
\begin{array}{ccc}
a_0 c_2 & 0 & b_2 \\
0 & a_0 & 0 \\
\frac{b_2}{a_0} & 0 & c_2
\end{array}
\right) \left(
\begin{array}{c}
t/\ell  \\
\phi  \\
\theta
\end{array}
\right) \,.
\label{J2 mixing}
\end{equation}
The nonvanishing parameters that define angular deficits are
\begin{eqnarray}
b_2 &=&a_0 \sinh \eta\,,  \notag\\
c_2 &=&\cosh \eta \,.\label{b2c2}
\end{eqnarray}
There are only two independent parameters, because $\left\{ a_0, b_2, c_2 \right\}$ satisfy
\begin{equation}
\det U=a_0^2 c_2^2 - b_2^2=a_0^2>0\,. \label{detU}
\end{equation}
The velocity parameter is related to the angular deficits as
$v=\frac{b_2}{a_0 c_2}$ with  $\left\vert v \right \vert <1$
(non-extremal brane), or $\left\vert b_2\right\vert <\left\vert
a_{0}c_{2}\right\vert$. The non-extremality condition does not mean
that extremal branes do not exist. It only means that they should be
constructed from a different Killing vector identification, not
continuously connected to the boost (\ref{boost-v}).

The periodicity in the angular coordinates of the brane geometry is
related to the magnitude of the identifications in the embedding
space,
\begin{equation}
\phi \simeq \phi +2\pi \quad \Leftrightarrow \quad \phi _{12}\simeq \phi_{12}+2\pi a_{0}\,,
\end{equation}
and
\begin{equation}
\begin{array}{ll}
\theta \simeq \theta +2\pi \quad \Leftrightarrow \quad & \phi_{05}\simeq \phi _{05}+2\pi b_{2}\,, \\ [0.5em]
& \phi _{34}\simeq \phi _{34}+2\pi c_{2}\,.
\end{array}
\label{two-ident}
\end{equation}
These correspond to the two linearly independent Killing vector
fields acting on $\mathbb{R}^{2,4}$, given by
\begin{eqnarray}
\xi_1 &=&2\pi a_{0}\,\mathbf{J}_{12}\,,  \notag \\ [0.4em]
\xi_2 &=&-2\pi b_2\,\mathbf{J}_{05}+2\pi c_2\,\mathbf{J}_{34}\,.
\label{two KV}
\end{eqnarray}
The identification by $\xi_1$ produces a static 2-brane in the
$x^1$-$x^2$-plane \cite{EGMZone}, and $\xi_2$ boosts it in the
perpendicular direction producing a spinning 2-brane if $b_{2}\neq
0$ (see Section \ref{ThreeD branes}). The corresponding norms are
\begin{eqnarray}
\left\Vert \xi_1 \right\Vert ^2 &=&4\pi^2 a_0^2B^2\,, \\ [0.5em]
\left\Vert \xi_2 \right\Vert ^2 &=&4\pi^2 A^2 \left( -b_2^2\cosh^2\!\tau +c_2^2\sinh^2\!\tau \right) \,.
\end{eqnarray}
The manifold formed after the identifications does not contain
closed time-like curves only if both $\xi_1$ and $\xi_2$ have
positive norms. While $\left\Vert \xi_1\right\Vert^2$ is always
nonnegative, $\left\Vert \xi_2\right\Vert^2$ is nonnegative only if
$c_2^2-b_2^2>0$, and
\begin{equation}
\tau \geq \tau_0=\tanh ^{-1}\frac{b_2}{c_2}\,.
\end{equation}
The resulting spacetime requires to cut off the region $\tau < \tau_0$.

%%%%%%%%%%%%%%%%%%%%%%%%% [3.3.1]
\subsubsection{Causal horizon}
%%%%%%%%%%%%%%%%%%%%%%%%%

In the spacetime with more than one Killing vector identification, we call the causal horizon a unique surface that separates the outer region, where \emph{all} $\left\Vert \xi_{i}\right\Vert ^{2}$ are strictly positive, from the interior region, where at least one of these norms is negative. Following the procedure established in previous sections, the interior must be removed from the manifold in order to avoid violations of causality produced by closed time-like curves. This surface should not be confused with an event horizon which can be crossed by a geodesic. The causal horizon becomes a boundary generated by the identification, where the spacetime ceases to be a smooth manifold once all points in this surface are identified and the surface shrinks the singular locus, $\left\Vert \xi_{i}\right\Vert^2=0$, where the brane sits. This transformation does not change the curvature of spacetime in the outer region, but it produces an infinite curvature at the horizon, that therefore becomes a brane source, $\mathbf{F}=\mathbf{j}$ \cite{EZ,MZtwo}.

In terms of the new coordinates, the metric (\ref{two spins}) reads
\begin{eqnarray}
ds^2 &=&\frac{\ell^2 dB^2}{A^2}+a_0^2 \,B^2 d\phi^2 + A^2d\tau^2 -\frac{A^2}{\ell^2 a_0^2} \left[\rule
{0pt}{14pt} b_2^2 -\left(b_2^2-a_0^4 c_2^2 \right) \cosh^2 \tau \right] \,dt^2  \notag \\
&&\!\!\!+A^2 \left[ \rule{0pt}{14pt}\left(c_2^2 -b_2^2\right) \cosh^2 \tau -c_2^2\right] d\theta ^2+\frac{2b_2 c_2}{\ell a_0} \, A^2 \left[ \rule{0pt}{14pt}\left( 1-a_0^2 \right) \cosh^2\!\tau -1\right] dt \,d\theta \,.  \label{J2 metric}
\end{eqnarray}
We shall argue now that this metric does not describe one codimension-two brane, but two intersecting codimension-two branes.

The first two terms in the metric (\ref{J2 metric}) describe a conical defect in the center of the $B\phi $-plane due to the angular deficit $2\pi \left( 1-a_0 \right)$. Thus, let us first analyze the center by setting $B=0$. This surface is a 2-brane,
\begin{eqnarray}
\left. ds^2\right\vert_{B=0} &=&-\left[\rule{0pt}{14pt}b_2^2-\left(b_2^2-a_0^4 c_2^2 \right) \cosh^2\!\tau \right] \,\frac{dt^2}{a_0^2} +\ell^2d\tau^2  \notag \\
&&+\ell^2 \left( \rule{0pt}{14pt}a_0^2 \cosh^2\!\tau -c_2^2\right) d\theta^2+\frac{2b_2 c_2\ell}{a_0}\,\left[ \rule{0pt}{14pt}\left(1-a_0^2\right) \cosh^2 \tau -1\right] dt\,d\theta \,.
\end{eqnarray}
Introducing a new radial coordinate, $\rho (\tau )$, that identifies the surface $\left\Vert \xi_2 \right\Vert =0$ with $\rho =0$, and also rescaling the time coordinate, as
\begin{equation}
\cosh \tau =\sqrt{\frac{c_2^2 + \frac{\rho^2}{\ell^2}}{c_2^2- b_2^2}}\,,\qquad T=\frac{a_{0}\,t}{c_2^2 - b_2^2}\,,\label{tau,T}
\end{equation}
leaves the metric in the ADM form,
\begin{equation}
\left. ds^2\right\vert_{B=0}=-\tilde{N}^2 dT^2+\frac{d\rho^2}{\tilde{N}^2}+\rho^2 \left(d\theta + \tilde{N}^{\theta} \, dT\right)^2\,, \label{3D-0brane}
\end{equation}
where
\begin{eqnarray}
& & \tilde{N}(\rho ) = \sqrt{b_2^2 + c_2^2+ \frac{\rho^2}{\ell^2} + \frac{\ell^2 b_2^2 c_2^2}{\rho^2}}\,, \label{lapse1}\\ [0.5em]
& & \tilde{N}^{\theta }(\rho) = \frac{b_2 c_2}{\ell a_0^2}\,\left(1- a_0^2\right) - \frac{b_2 c_2\ell}{\rho^2}\,. \label{shift1}
\end{eqnarray}
Comparing with Eqs.(\ref{Schwarzschild-like}) and (\ref{lapse-shift}), the worldvolume geometry (\ref{3D-0brane}) describes the
three-dimensional spacetime produced by a spinning 0-brane in anti-de Sitter. This is the solution, for example, of  Einstein's equations in three dimensions with negative cosmological constant in the presence of a spinning topological defect. The mass of the brane is $M=-(b_2^2+ c_2^2)<0$ and $J=2\ell c_2b_2$ is the angular momentum, and thus the familiar inequality $\ell \left\vert M\right\vert >\left\vert J\right\vert$ follows trivially.

In general, for arbitrary $B$, the full five-dimensional metric describes a warped product of a conical defect and a spinning brane,
\begin{equation}
ds^2= \frac{\ell^2\, dB^2}{B^2+\ell^2}+ a_0^2\,B^2 d\phi^2+\frac{B^2+ \ell^2}{\ell^2}\left. ds^2\right\vert _{B=0}\,.
\end{equation}
Thus, using the coordinates $T$ and $\rho$ defined above, and introducing $r=a_{0}B$, the metric reads
\begin{equation}
ds^2=\frac{dr^2}{f^2(r)}+ r^2 d\phi^2+\frac{f^2(r)}{a_0^2} \left[-\tilde{N}^2dT^2 + \frac{d\rho^2}{\tilde{N}^2}+ \rho^2 \left(d\theta+ \tilde{N}^{\theta }\,dT\right)^2\right] \,,
\label{interbrane}
\end{equation}
where
\begin{equation}
f^2(r)=a_0^2+\frac{r^2}{\ell^2}\,.
\end{equation}
Note that the lapse and shift functions defined in Eq.(\ref{ADM angles}), here correspond to $N=f\tilde{N}/a_0$ and $N^\theta =\tilde{N}^\theta$, written in the coordinates $(r,\rho, T, \phi, \theta)$ that cover only the outer region with respect to the causal horizon.

It is apparent now that the surface $r=0$ is a three-dimensional
submanifold with the geometry produced by a spinning 0-brane. On the
other hand, the surface $\rho=0$ (where $\theta$ is not defined), has the geometry of a static 2-brane.
The resulting warped geometry (\ref{interbrane}) is such that the
singularities representing the worldvolumes of the branes are, in
turn, the 3D spacetime geometries produced by a static ($\rho=0$)
and a spinning ($r=0$) 0-branes, respectively.

This suggests that this configuration might survive the addition of a further interaction corresponding to codimension-four objects localized in the worldvolume of both 2-branes. We will explore this possibility elsewhere.

We expect that the form of the sources keeps this geometrical picture of two intersecting 2-branes.

%%%%%%%%%%%%%%%%%%%%%%%%%%%%%%%%%%&&% [3.3.2]
\subsubsection{Sources for intersecting 2-branes}
%%%%%%%%%%%%%%%%%%%%%%%%%%%%%%%%%%&&%

Following the steps of section 3.2.1 to identify the source, let us consider a generic Lie group element (\ref{g general}) in the coordinates (\ref{J2 mixing}),
\begin{equation}
g(t,\phi ,\theta )=g_0\,e^{-\left( \frac{b_2t}{a_0\ell }+c_2\theta \right) \mathbf{J}_{34}}e^{\left( \frac{a_0 c_2 t}{\ell}+ b_2\theta\right) \mathbf{J}_{05}}\,e^{-a_0\phi \mathbf{J}_{12}}\,e^{\tau \mathbf{J}_{35}}e^{p(B)\mathbf{J}_{15}}\,,
\end{equation}
where $g_0$ is some constant group element. Similarly to the three-dimensional case, the curvature singularity can be calculated evaluating the loop integral of $g^{-1}dg$ along an infinitesimal contour around the Killing vector horizons. In this case, the source is identified as
\begin{eqnarray}
\mathbf{j} &=&-2\pi a_0\,\mathbf{J}_{12}\,\delta (\Sigma _{12}) + \frac{2\pi b_2}{\ell}\,\left( A\cosh\tau\, \mathbf{J}_{05}+B\cosh\tau \,\mathbf{J}_{01}+\ell \sinh\tau \,\mathbf{J}_{03}\right) \delta(\Sigma_{05})  \notag \\ [0.5em]
&&+\frac{2\pi c_2}{\ell} \left(A\sinh\tau\, \mathbf{J}_{45}-B\sinh\tau \,\mathbf{J}_{14}-\ell \cosh\tau\, \mathbf{J}_{34}\right)
\delta(\Sigma_{34}) \,.
\label{J_intersect}
\end{eqnarray}
Since $\mathbf{j}$ is a sum of various currents whose Dirac deltas are supported on (at most two) different submanifolds of the four-dimensional spatial section, it means that we are dealing with (at most two) different independent 2-branes. Indeed, the first term in the source corresponds to a static 2-brane,
\begin{equation}
\mathbf{j}_{1}^{\text{static}}=-2\pi a_0\,\delta
(\Sigma_{12})\,,\qquad \delta (\Sigma_{12})=\frac{1}{2\pi
}\,\delta(r)\,dr\wedge d\phi \,.
\end{equation}
Expressing $\delta (\Sigma_{05})$ and $\delta (\Sigma_{34})$ in terms of $\delta (\rho)$, using the relation for $\tau(\rho)$ given by Eq.(\ref{tau,T}) and finally evaluating the current by means of the identity $\rho \delta (\rho)=0$, it can be shown that
the source for the 3D spinning brane is,
\begin{equation}
\mathbf{j}_2^{(3D\text{ spinning})}=\frac{2\pi}{\sqrt{c_2^2-b_2^2}}
\left[-c_2^2\mathbf{J}_{34}+ b_2^2\mathbf{J}_{03}+b_2c_2 \left(
\mathbf{J}_{05}+\mathbf{J}_{45}\right) \right]  \delta (\rho)\,d\rho
\wedge \frac{d\theta}{2\pi }\,.
\end{equation}
This can be confirmed by direct comparison with Eq.(\ref{j_holonomy}). We conclude that an intersection of two 2-branes is a 0-brane (that is, a worldline parameterized by the time $T$) with nonvanishing angular momentum. In general, when $\rho \neq 0$, these two branes combine the sources in a more complicated way, as in Eq.(\ref{J_intersect}).

We shall see below that intersecting codimension-two branes \emph{generically} appear when the angular momentum of the solution is transversal to the original static brane.

%%%%%%%%%%%%%%%%%%%%%%%%%%%%%%%%%%%%%%%%%%%%%%%%%%%% [3.4]
\subsection{2-branes with two angular momenta} \label{5Dbrane_two J}
%%%%%%%%%%%%%%%%%%%%%%%%%%%%%%%%%%%%%%%%%%%%%%%%%%%%

In order to boost the static 2-brane in a most general way in five dimensions, we introduce two Lorentz transformations with velocities $w=\tanh \zeta$ and $v=\tanh \eta$ ($0\leq v$,$w<1$),
\begin{eqnarray}
e^{-\zeta\mathbf{L}_{02}} &:&\qquad t'=\frac{t_{0}-\ell
w\phi _{0}}{\sqrt{1-w^{2}}}\,,\qquad \phi =\frac{\phi _{0}-\frac{w}{\ell} t_0}{\sqrt{1-w^{2}}}\,, \\
e^{-\eta\mathbf{L}_{04}} &:&\qquad t=\frac{t'-\ell v\theta_0}{\sqrt{1-v^2}}\,,\qquad \theta =\frac{\theta_0-\frac{v}{\ell}\,t'}{\sqrt{1-v^2}}\,,
\end{eqnarray}
represented by matrices (\ref{Lambda1}) and (\ref{Lambda2}), where $\left[\mathbf{L}_{02},\mathbf{L}_{04}\right] =\mathbf{L}_{24}$. After that double boosting, the mixing of angles given by $U=U_0\, e^{\zeta\mathbf{L}_{02}} e^{\eta\mathbf{L}_{04}}$ takes the form
\begin{equation}
U=\left(
\begin{array}{ccc}
a_{0}\cosh \zeta\cosh \eta & a_0\sinh \zeta & a_0\cosh \zeta \sinh \eta \\
a_{0}\sinh \zeta\cosh \eta & a_0 \cosh \zeta & a_0 \sinh \zeta \sinh \eta \\
\sinh \eta & 0 & \cosh \eta
\end{array}
\right) \,.
\end{equation}

From $\phi \simeq \phi +2\pi $ and $\theta \simeq \theta +2\pi $, we get that angular momenta are produced by two independent identifications,
\begin{equation}
\begin{array}{ll}
\phi_{12}\simeq & \phi_{12}+2\pi a_{1}\,, \\
\phi_{05}\simeq & \phi_{05}+2\pi b_{1}\,, \\
\phi_{34}\simeq & \phi_{34}\,,
\end{array}
\label{identifications1}
\end{equation}
and
\begin{equation}
\begin{array}{ll}
\phi _{12}\simeq & \phi _{12}+2\pi a_{2}\,, \\
\phi _{05}\simeq & \phi _{05}+2\pi b_{2}\,, \\
\phi _{34}\simeq & \phi _{34}+2\pi c_{2}\,,
\end{array}
\label{identifications2}
\end{equation}
respectively, where the constants are
\begin{equation}
\begin{array}{llll}
a_1= & a_0\cosh\zeta , & a_2= & a_0\sinh\zeta \sinh\eta , \\
b_1= & a_0\sinh\zeta ,\qquad & b_2= & a_0\cosh\zeta \sinh\eta , \\
c_1= & 0\,, & c_2= & \cosh\eta .\label{constants}
\end{array}
\end{equation}
There are only three independent parameters, because six constants are subjected to three constraints,
\begin{equation}
a_1^2-b_1^2=a_0^2,\qquad a_1 a_2=b_1 b_2,\qquad a_1^2\left( c_2^2-1\right) =b_2^2\,. \label{constants constr}
\end{equation}

The identifications (\ref{identifications1}) and (\ref{identifications2}) are produced by two independent Killing vector fields
\begin{eqnarray}
\xi_1 &=&2\pi a_1 \mathbf{J}_{12}-2\pi b_1\,\mathbf{J}_{05}, \notag \\
\xi_2 &=&2\pi a_2 \mathbf{J}_{12}-2\pi b_2\,\mathbf{J}_{05}+2\pi c_2\mathbf{J}_{34}\,,
\end{eqnarray}
whose norms read
\begin{eqnarray}
\left\Vert \xi_1\right\Vert^2 &=&4\pi^2\left(a_1^2 B^2-b_1^2 A^2\cosh^2\!\tau \right) \,,  \notag \\
\left\Vert \xi_2\right\Vert^2 &=&4\pi^2 \left(a_2^2B^2-b_2^2 A^2\cosh^2\!\tau +c_2^2 A^2\sinh^2\!\tau \right) \,.
\end{eqnarray}
The vector $\xi_1$ is the same one that brings into existence a 2-brane with the angular momentum $J_\phi$, and we know that its norm is positive when Eqs. (\ref{xi_1>0}) and (\ref{B_h}) are satisfied, that is,
\begin{equation}
B\geq B_1(\tau )\,,\qquad \tau <\tau _{01}\,,
\end{equation}
where
\begin{eqnarray}
B_1(\tau ) \equiv \frac{b_1\ell \cosh \tau }{\sqrt{a_1^2-b_1^2 \cosh^2\!\tau }}\,, \qquad
\tau_{01} \equiv \cosh^{-1}\frac{a_1}{b_1}\,.
\end{eqnarray}
On the other hand, sectors with $\left\Vert \xi_2\right\Vert^2 \geq 0$ exist only if $c_2^2-b_2^2>0$. Then $\xi_2$ is a space-like vector in the region
\begin{equation}
B\geq B_2(\tau ) \equiv \left\{
\begin{array}{ll}
\ell \sqrt{\left\vert \frac{b_2^2-\left(c_2^2-b_2^2\right) \sinh^2\!\tau }{a_2^2-b_2^2+ \left(c_2^2-b_2^2\right) \sinh^2\!\tau}\right\vert }\,,\quad & \text{if }\tau_{02}<\tau \leq \tau_{03}\,, \\
0\text{\thinspace }, & \tau >\tau_{03}\,,
\end{array}
\right.  \label{B>}
\end{equation}
where we have introduced the points
\begin{eqnarray}
\tau_{02} =\sinh^{-1}\sqrt{\left\vert \frac{b_2^2-a_2^2}{c_2^2-b_2^2}\right\vert }\,,\qquad
\tau_{03} =\sinh^{-1}\frac{b_2}{\sqrt{\left\vert c_2^2-b_2^2\right\vert }}\geq \tau_{02}\,.
\end{eqnarray}

\begin{figure}[h]
\centering
\includegraphics[width=0.43\textwidth]{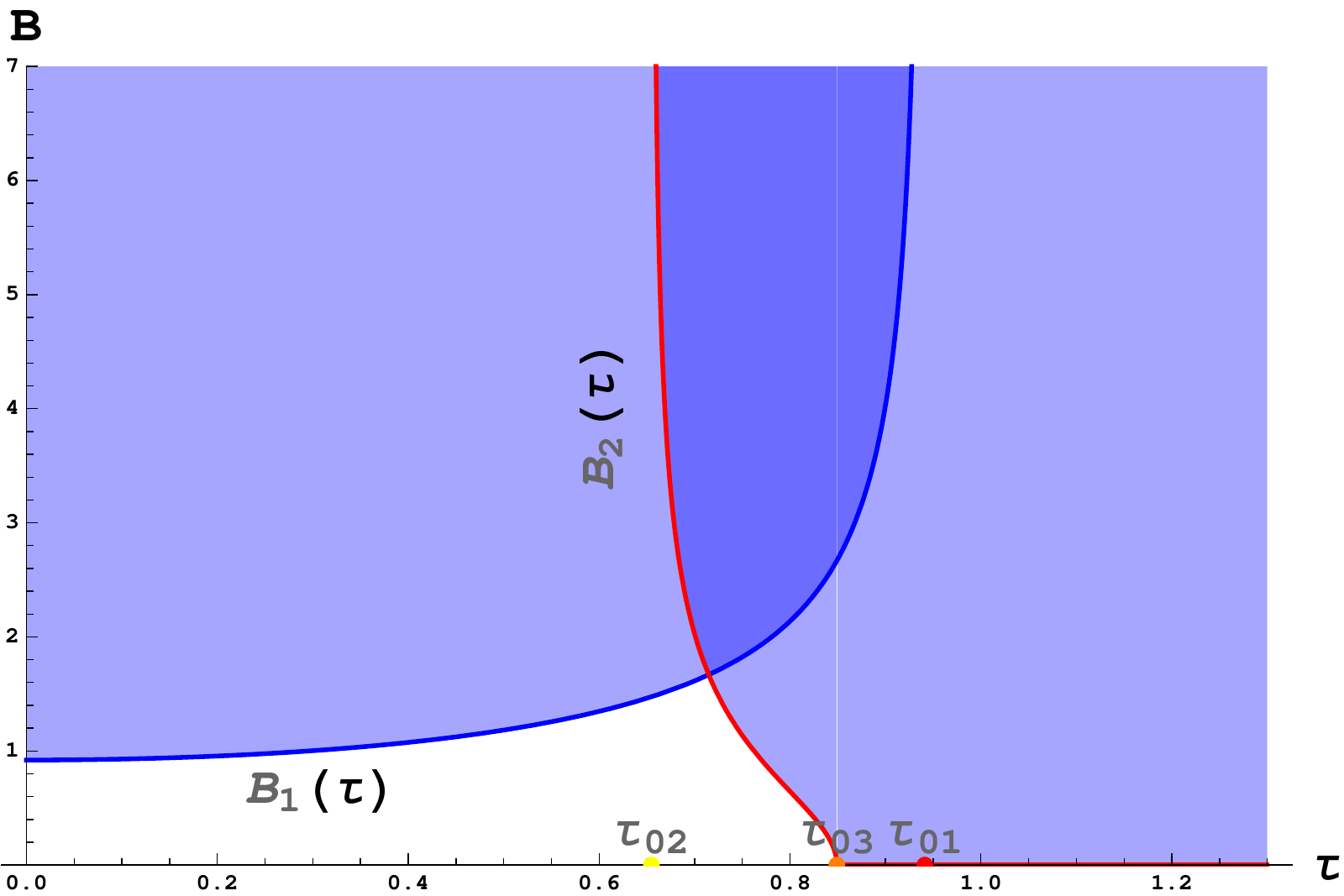} \quad
\includegraphics[width=0.43\textwidth]{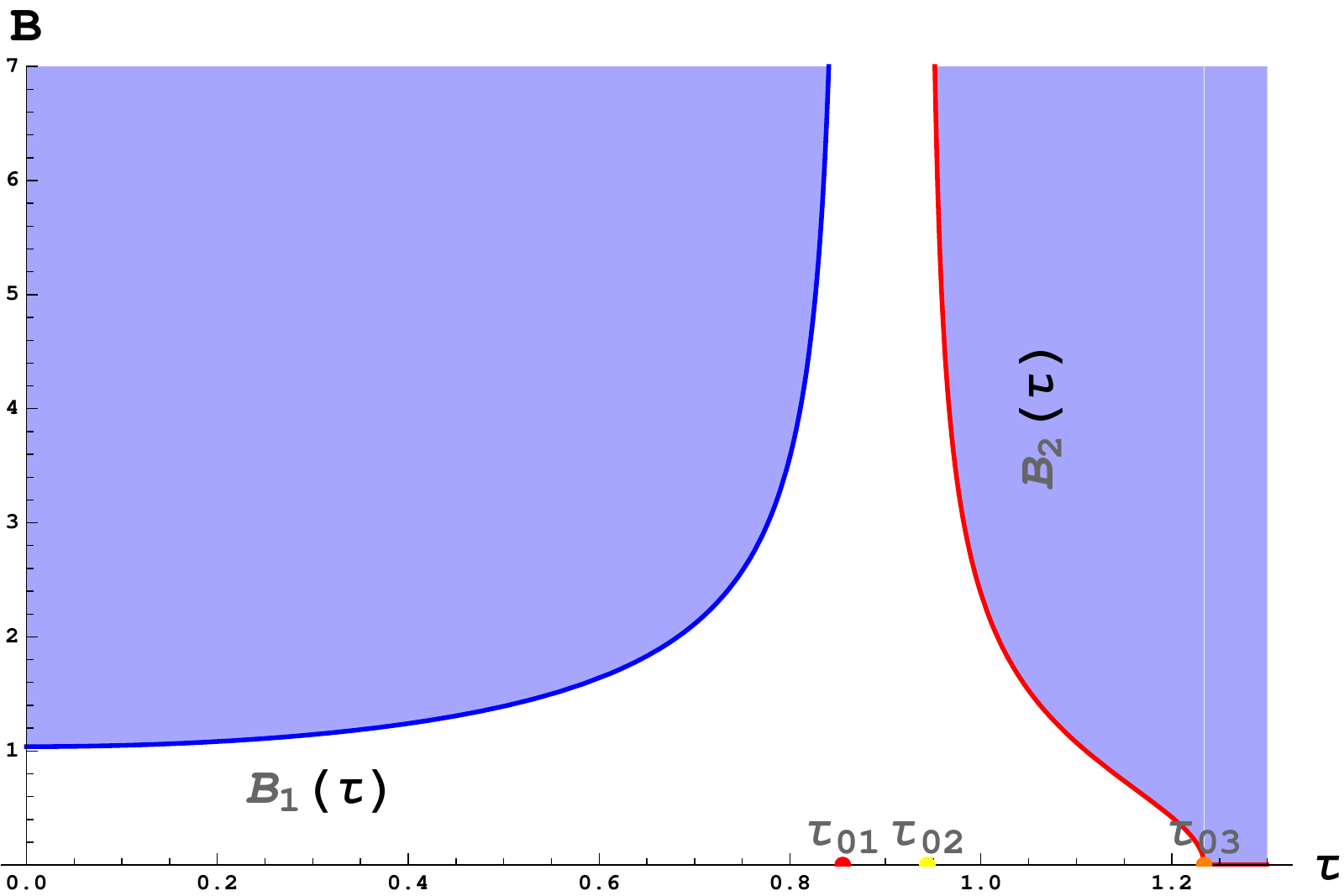}
\caption{Two cases can be distinguished according to the relative position of $\tau_{01}$ and $\tau_{02}$.}
\label{5Dbrane-two}
\end{figure}

When $\tau =\tau_{03}=\tau_{02}$, the vector $\xi_2$ is space-like for any $B$, but this 2-brane possesses one angular momentum only, since $a_2=0$. If $c_2^2-b_2^2<0$, the vector $\xi_2$ is not space-like in any region of the manifold. Regions where a 2-brane is defined for various angular momenta, are shown in Figure \ref{5Dbrane-two}.

%%%%%%%%%%%%%
%%%%%%%%%%%%%

Finally, if $\zeta$, $\eta\neq 0$, a well-defined spacetime corresponds to a region where both $\xi _{1}$ and $\xi _{2}$ are space-like vectors, when the horizons (with vanishing $\left\Vert \xi_1\right\Vert$ and $\left\Vert \xi_2\right\Vert $) are identified with a point. In this procedure, all the sectors where a norm of at least one of the vectors is negative, have been removed. What remains is non empty if the curves $B_1(\tau )$ and $B_2(\tau )$ intersect, that is, if $c_2^2-b_2^2>0$ and $\tau_{01}\geq \tau_{02}$. The intersection point $\tau_c$ is a solution of $B_1(\tau_c)=B_2(\tau_c)$. The causal horizon $B_*(\tau )$ is the boundary of the region where both Killing vectors are space-like, so that it coincides in parts with the curves $B_1(\tau )$ and $B_2(\tau )$. In terms of the parameters, a condition to have a 2-brane is
\begin{equation}
c_2^2-b_2^2>0\,,\qquad b_1^2\left( b_2^2-a_2^2\right) \leq a_0^2\left( c_2^2-b_2^2\right) \,. \label{existence}
\end{equation}

As mentioned above, when the intersection exists, the causal horizon $B_*(\tau ),\;\tau _{02}\leq \tau \leq \tau _{01}$ is the union of two surfaces defined by $B_1(\tau),\;\tau_c \leq \tau \leq \tau_1$ and $B_2(\tau),\;\tau_{02}\leq \tau \leq \tau_c$. The radial coordinate outside the brane is defined for $r=B-B_*(\tau)\geq 0$, and the radial coordinate $\tau$ inside the brane becomes compact, $\tau_{02}\leq \tau \leq \tau_{01}$.

Removing the interior region ($B-B_*(\tau)<0$) produces a globally nontrivial identification. Unlike identifications generated by the action of a Killing vector, this one \textit{does} change the curvature, but only at the horizon, whereas for $B>B_{\ast}(\tau)$ (outside the horizon) the equation $\mathbf{F}=0$ holds, and the spacetime is locally AdS. On the other hand, the horizon $B_{\ast}$ becomes a point after the identifications, and the curvature there is infinite.

The surfaces $B_1$ and $B_2$ are not joined smoothly at the point $\tau_c$ because, as in Section \ref{Intersection}, this point is an intersection of two 2-branes, produced by two distinct Killing vector identifications. This can be best seen directly from the source. When the conditions (\ref{existence}) are fulfilled and the branes exist, the curvature singularity is calculated from
\begin{eqnarray}
\oint g^{-1}dg &=&\frac{1}{\ell }\,\left( B_*\mathbf{J}_{25}-A_*\mathbf{J}_{12}\right) \oint d\phi _{12}  \notag \\
&&+\frac{1}{\ell }\,\left[ \cosh \tau \left( A_*\mathbf{J}_{05}+B_*\mathbf{J}_{01}\right) +\ell \sinh \tau \,\mathbf{J}_{03} \right] \oint d\phi _{05}  \notag \\
&&+\frac{1}{\ell }\,\left[ \sinh \tau \left( A_*\mathbf{J}_{45}-B_*\mathbf{J}_{14}\right) -\ell \cosh \tau \,
\mathbf{J}_{34}\right] \oint d\phi _{34}\,.
\end{eqnarray}
The integral is taken around an infinitesimal path close to the horizon $B_\ast$. Note that this path is not smooth, since it is composed by two curves.

Non-vanishing integrals are due to the presence of sources, and the total current is a sum of various sources defined on different planes. Again, these sources are expressed in terms of only two independent Dirac deltas -- the ones on the $B$-$\phi$ and $\tau$-$\theta$ planes, generated by two independent Killing vector identifications. This is guaranteed by construction because the sources appear only at the causal horizons $\left\Vert \xi_i\right\Vert=0$.

Because $g_{\theta\phi} \neq 0$, the angular section $\gamma_{mn}$ in the ADM metric (\ref{ADM angles}) is not diagonal,  the $B$-$\phi$ and $\tau$-$\theta$ planes are not orthogonal to each other. Thus, the coordinates used here are natural to perform the identifications, but they are not the best adapted for writing the sources for each brane as an independent term. A convenient method to deal with this is the one used in Ref.\cite{Harmark} for locally flat configurations, where the metric is put in the canonical form that separates (up to warp factors) submanifolds with different Killing vector symmetries. This calculation is technically cumbersome because the coordinate transformation depends also on $B$ and $\tau$, and is beyond the scope of this paper.

%%%%%%%%%%%%%%%%%%%%%%%%%%%%%%%%%%%%%%%%%%%%%%%%% [3.4.1]
\subsubsection{Inequivalent 2-branes with two angular momenta}
%%%%%%%%%%%%%%%%%%%%%%%%%%%%%%%%%%%%%%%%%%%%%%%%

Since Lorentz transformations do not commute, one last question to clarify is what happens if the order of the boosts applied to a static brane is reversed. Taking $U=U_0e^{\eta \mathbf{L}_{04}}e^{\zeta\mathbf{L}_{02}}$, an inequivalent mixing of the angles is obtained, so that the parameters for the angular deficits now become
\begin{equation}
\begin{array}{llll}
a_1= & a_0\cosh \zeta \medskip & a_2= & 0\,, \\
b_1= & a_0\sinh \zeta \cosh \eta ,\medskip \qquad & b_2= & a_0\sinh \eta , \\
c_1= & \sinh \zeta\sinh \eta , & c_2= & \cosh \eta.
\end{array}
\end{equation}
The constraints between the parameters are
\begin{equation}
a_0^2c_1c_2=b_1b_2\,,\qquad a_1^2-\frac{b_1^2}{c_2^2} =a_0^2\,,\qquad b_2^2=a_0^2(c_2^2-1)\,.
\end{equation}

\begin{figure}[h]
\centering
\includegraphics[width=0.43\textwidth]{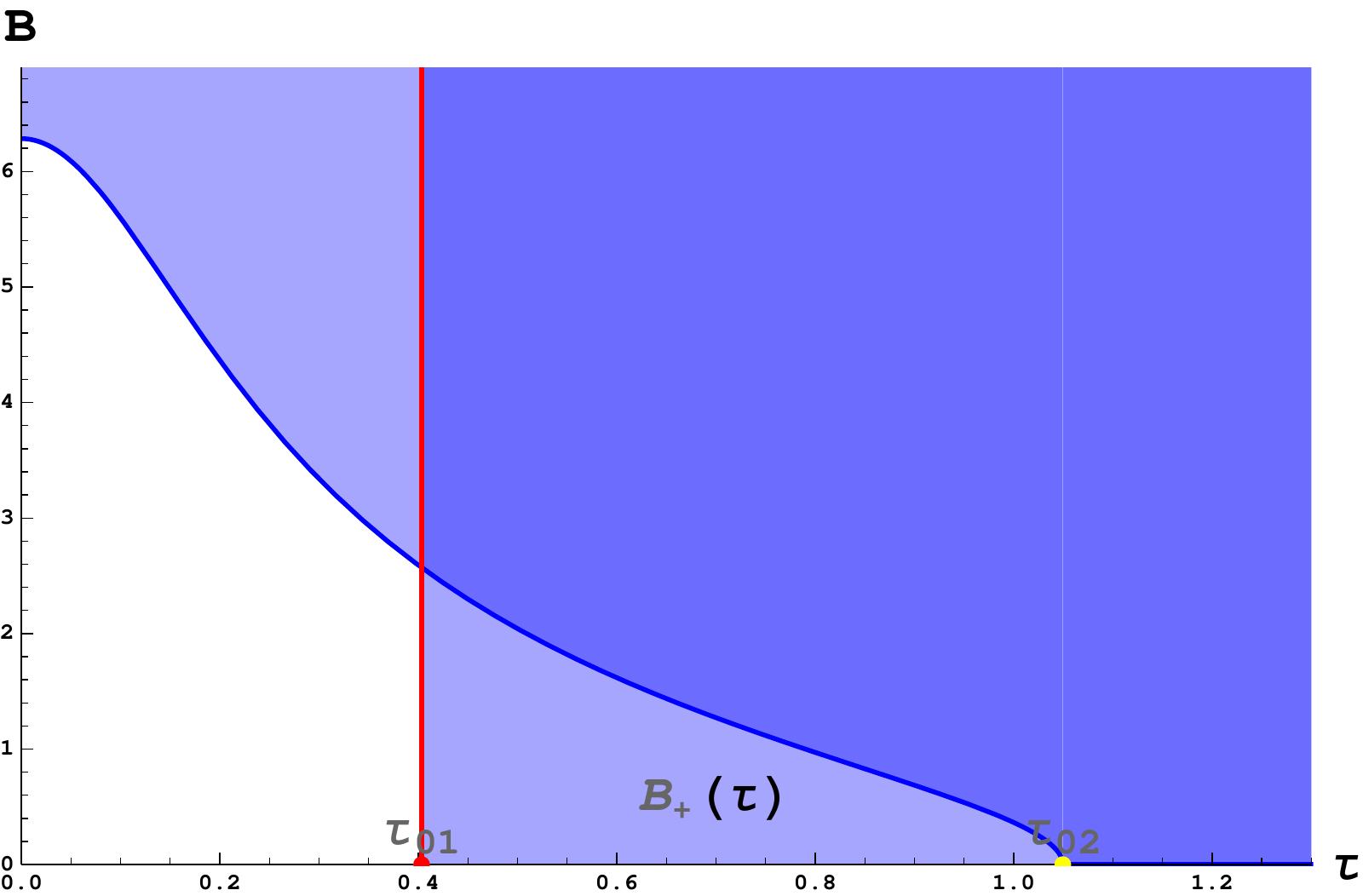} \quad
\includegraphics[width=0.43\textwidth]{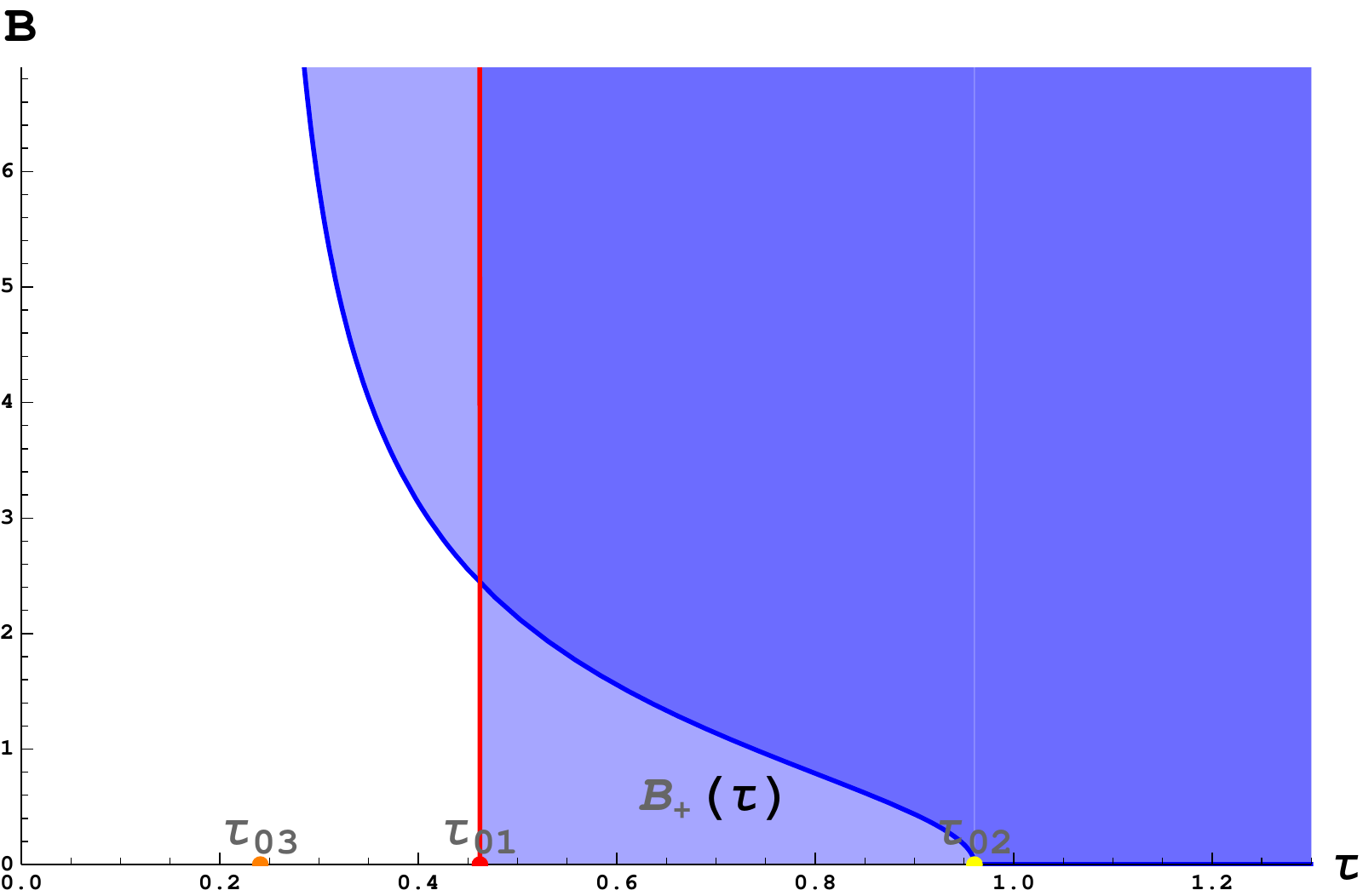} \quad
\includegraphics[width=0.43\textwidth]{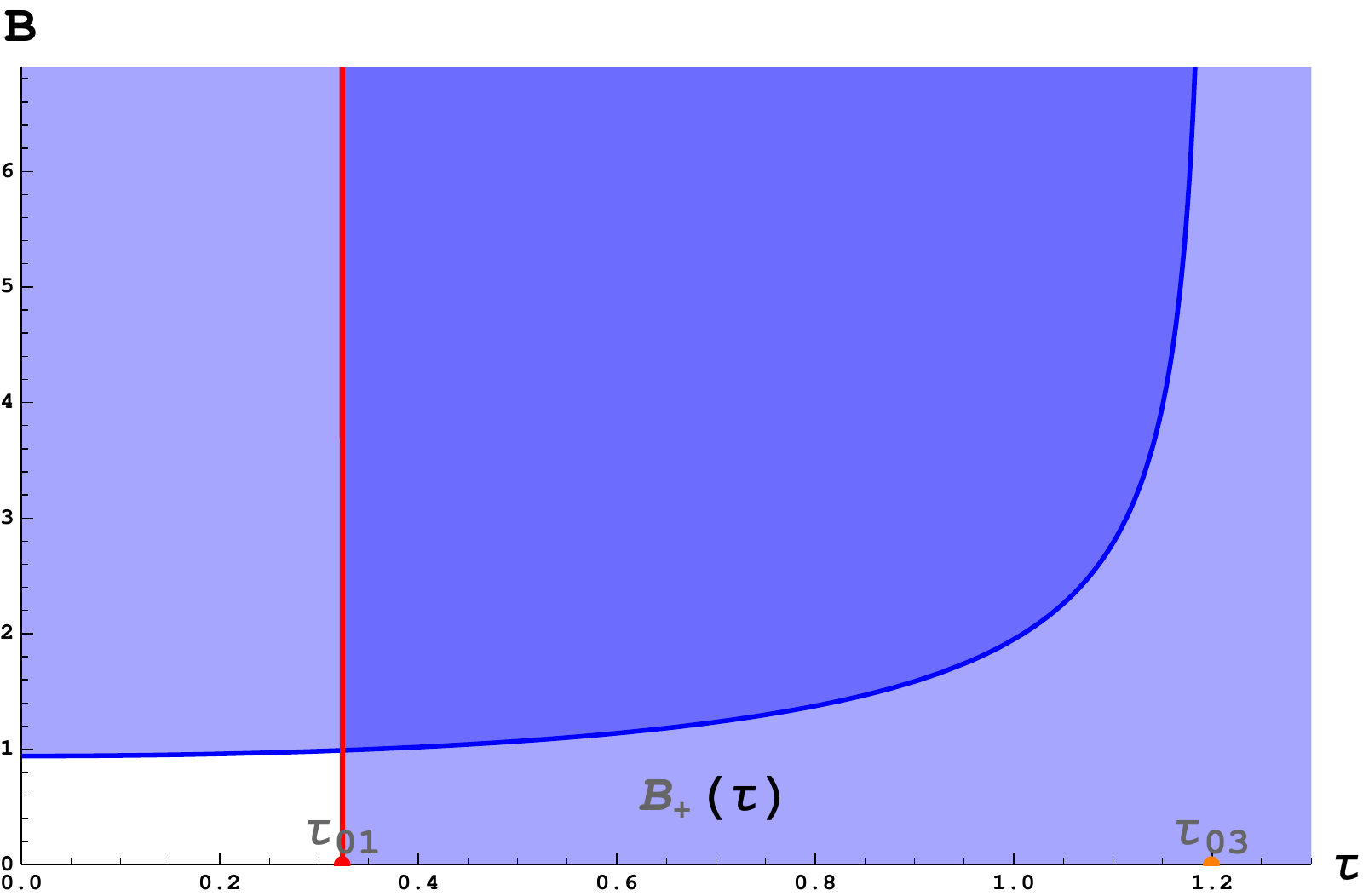}
\caption{The doubly shaded area represents the regions where both Killing vectors are space-like. Three cases can be distinguished depending on the relative position of $\tau_{01}$ with respect to the pair defined by $\tau_{02}$ and $\tau_{03}$ (whenever they adopt real values).}
\label{5Dbrane-two-bis}
\end{figure}

The norms of Killing vectors that produce identification read
\begin{eqnarray}
\left\Vert \xi_1\right\Vert^2 &=&4\pi^2\left(a_1^2 B^2-b_1^2 A^2\cosh^2\!\tau +c_1^2 A^2\sinh^2\!\tau \right) \,, \notag \\
\left\Vert \xi_2\right\Vert^2 &=&4\pi ^2 A^2\left(-b_2^2 \cosh^2\!\tau +c_2^2 \sinh^2\!\tau \right) \,.
\end{eqnarray}
The vector field $\xi_2$ is space-like for $\tau \geq \tau _{01}$, where $\tau _{01}= \sinh^{-1} \frac{b_2}{\sqrt{c_2^2 - b_2^2}}$ exists if $c_2^2 - b_2^2 > 0$, otherwise $\xi_2$ is never space-like, in which case the spacetime would have ill-defined causal structure, making those brane solutions unphysical. On the other hand, if $c_1^2-b_1^2>0$, the norm of the $\xi_1$ can be written as
\begin{equation}
\left\Vert \xi_1\right\Vert^2 = 4\pi^2 ( c_1^2-b_1^2) \left[ B^2  ( \sinh^2\tau -\sinh^2\tau_{03}) +\ell^2 (\sinh^2\tau -\sinh^2\tau_{02})\right] \,,
\end{equation}
where $\tau _{02}=\sinh^{-1}\frac{b_1}{\sqrt{c_1^2 - b_1^2}}$ and $\tau _{03}=\sinh^{-1}\sqrt{\frac{b_1^2-a_1^2}{c_1^2 - b_1^2}}$. Then, the regions with nonegative $\left\Vert \xi_1\right\Vert^2$ are shown in Figure \ref{5Dbrane-two-bis}, and they are bounded by the horizon $B_+(\tau)$.

%%%%%%%%%%%%%
%%%%%%%%%%%%%

A nonempty intersection of two regions (where both $\xi_1$ and $\xi_2$ are spacelike) is what remains after all other sectors are cut out from the manifold. This truncation produces the sources, that geometrically describe two intersecting branes.

Compared to the two 2-branes of Subsection \ref{5Dbrane_two J} obtained by different boosting of the static brane, the configurations constructed here contain branes with noncompact internal spacetimes ($B$ and $\tau$ unbounded), as shown in the first two graphs of  Figure \ref{5Dbrane-two-bis}.

Note that for a 2-brane whose angular momenta are defined by the boost $e^{\zeta\mathbf{L}_{02}} e^{\eta\mathbf{L}_{04}}$, the $U$ matrix can be written as
\begin{equation}
U=U_0\,e^{\alpha \mathbf{L}_{02}+\beta \mathbf{L}_{04}+\gamma \mathbf{L}_{24}}\,,  \label{U_gen}
\end{equation}
where the parameters are obtained using the Baker-Campbell-Hausdorff formula as $\alpha =\frac{\zeta}{\eta}\sin\eta$, $\beta =\eta$ and $\gamma =\frac{\zeta}{\eta}(1-\cos\eta)$. Thus, in general, a 2-brane with the required features can be  ``designed" by appropriately choosing the parameters $\alpha$, $\beta $ and $\gamma$. In order to ensure that $U$ corresponds to a boost and not a rotation between the coordinates $\phi$ and $\theta$,  the generator should have real eigenvalues. The eigenvalues of $U$ squared are
\begin{equation}
z^2=\alpha^2+\beta^2-\gamma^2>0\,,
\end{equation}
which can are, in turn, related to the angular deficits in $\phi$ and $\theta$ by
\begin{eqnarray}
a_1 &=&\frac{a_0}{z^2}\,\left[\beta ^2+\left(\alpha ^2-\gamma ^2\right) \cosh z\right] \,,  \notag \\
a_2 &=&\frac{a_0}{z^2}\,\left[\gamma z\sinh z+\alpha \beta \left(\cosh z-1\right) \right] \,,  \notag \\
b_1 &=&\frac{a_0}{z^2}\,\left[\alpha z\sinh z+\beta \gamma \left(\cosh z-1\right) \right] \,,  \notag \\
b_2 &=&\frac{a_0}{z^2}\,\left[\beta z\sinh z+\alpha \gamma \left(\cosh z-1\right) \right] \,,  \notag \\
c_1 &=&\frac{1}{z^2}\,\left[-\gamma z\sinh z+\alpha \beta \left(\cosh z-1\right) \right] \,,  \notag \\
c_2 &=&\frac{1}{z^{2}}\,\left[\alpha^2+\left(\beta ^2-\gamma ^2\right) \cosh z\right] \,.
\end{eqnarray}

%%%%%%%%%%%%%%%%%%%%%%%%%%%%%%%%%%%%%% [3.5]
\subsection{BPS spinning 2-brane configurations}
\label{BPS in 5D}
%%%%%%%%%%%%%%%%%%%%%%%%%%%%%%%%%%%%%%%

In order to investigate the stability of these $p$-branes, one could
try to analyze the dynamics of their perturbations. This requires a
previous knowledge of the Lagrangian that we have not specified so
far. We will study stability in the context of CS AdS supergravity
for the super group $SU(2,2|N)$, which is the natural supersymmetric
extension of the five-dimensional AdS group. This supergravity has
$N$ supersymmetries and the spinors $\psi _{s}$ ($s=1,\ldots ,N$)
transform as $SU(N)$ vectors.

A $p$-brane is stable if it preserves some supersymmetries described by local spinorial parameters $\epsilon_s(x)$. These Killing spinors are solutions of
\begin{equation}
D(\mathbf{A}) \epsilon_s =0,
\label{k-spinor-eq}
\end{equation}
in the background given by the AdS connection (\ref{AdS connection}). Hence, stability will be expected to hold for those configurations that admit globally defined solutions of (\ref{k-spinor-eq}). For simplicity, we will assume that all other components of $\mathbf{A}$ in the superalgebra vanish. The AdS generators $\mathbf{J}_{AB}$ are represented in terms the five-dimensional Gamma matrices, $\Gamma_a$, as $\mathbf{J}_{ab}=\frac{1}{4}\,[\Gamma_a,\Gamma_b]$ and $\mathbf{J}_{a5}=\frac{1}{2}\,\Gamma_a$.

Solutions to this equation have been discussed in detail for static 2-branes in Ref.\cite{EGMZone}. Here, we make use of the fact that a solution for spinning 2-branes has the same local form as the static solution, where only global (boundary) conditions are changed. Thus, we write a general solution for the Killing spinor as
\begin{equation}
\epsilon _{s}=e^{-\frac{1}{2}\,p(B)\Gamma _{1}}e^{-\frac{1}{2}\,\tau \Gamma _{3}}e^{\frac{i}{2}\,\left( \phi _{12}+\phi _{34}-\phi _{05}\right) }\,\chi_s\,,
\end{equation}
where $p(B)=\sinh^{-1}(B/\ell)$, and a constant common eigen-spinor $\chi_s$ of the commuting matrices satisfies
\begin{equation}
\Gamma _{0}\,\chi_s=i\chi_s\,,\qquad \Gamma _{1}\Gamma _{2}\,\chi_s=i\chi_s\,,\qquad \Gamma _{3}\Gamma _{4}\,\chi_s=i\chi_s\,.
\label{eta condition}
\end{equation}
Only the last two conditions are independent because $\Gamma _{0}=-i\Gamma _{1}\Gamma _{2}\Gamma_{3}\Gamma _{4}$. There are, thus, a priori $N/4$ independent spinors $\epsilon_s$.

The existence of nontrivial solutions for (\ref{k-spinor-eq}) depends on the boundary conditions  of $\epsilon_s(\phi,\theta)$. The periodicity of coordinates $\phi,\theta$ are determined by the mapping (\ref{mix matrix}) between these angles and the embedding angles. (Anti)periodic conditions $\epsilon_s(\phi +2\pi )=\pm \epsilon_s(\phi )$ and $\epsilon_s(\theta +2\pi) = \pm \epsilon_s(\theta )$ imply
\begin{equation}
a_{i}-b_{i}+c_{i}=n_{i}\in \mathbb{Z}\,,\qquad (i=1,2)\,.
\label{double extremal}
\end{equation}
This establishes the BPS condition that relates the parameters $a_i,\;b_i$ and $c_i$, which are the same for all branes with equal  angular deficits, regardless of the boosts by which they are spun.  In general, these are not the same as the extremality  conditions that match the mass and angular momentum parameters for the extremal branes that correspond to the boosted static brane with maximal velocity. Only in three dimensions these conditions coincide.

Clearly, in the absence of angular deficits ($a_i,b_i,c_i\in \mathbb{Z}$) there exists one stable solution, namely, global AdS. For a static 2-brane, there is only one nontrivial parameter, which is not an integer. Hence the extremality conditions (\ref{double extremal}) cannot be fulfilled --the static 2-brane without  bosonic CS matter is not stable \cite{EGMZone}. If there is a nonvaninshing angular momentum, then two or more parameters are not integers, and the extremality conditions (\ref{double extremal}) could in principle be satisfied by a one-parameter family of configurations. However, fulfilment of this condition is not guaranteed in general and must be checked for each particular system of 2-branes. We, therefore, turn to particular cases.

For intersecting 2-branes with one angular momentum $J_\theta$, the independent nontrivial parameters are $b_2$ and $c_2$, given by Eq.(\ref{b2c2}), and there is a unique extremality condition
\begin{equation}
c_2-b_2=n_2\,.
\end{equation}
This condition clearly constraints only one parameter while leaving the others arbitrary, and is consistent with (\ref{detU}), so these stable BPS branes exist.

Similarly, for a massive 2-brane with angular momentum $J_\phi$, two independent parameters are $a_{1}$ and $b_{1}=\sqrt{a_{1}^{2}-a_{0}^{2}}$, and a stable BPS 2-brane is obtained when
\begin{equation}
a_{1}-\sqrt{a_{1}^{2}-a_{0}^{2}}=n_{1}\,.
\end{equation}

More generally, for intersecting 2-branes with two angular momenta $J_\phi$ and $J_\theta$, whose angular deficits have the form (\ref{constants}), the extremality condition reads
\begin{eqnarray}
a_{0}\left( \cosh \eta _{1}-\sinh \eta _{1}\right)  &=&n_{1}\,,  \notag \\
\cosh \eta _{2}-a_{0}n_{1}\sinh \eta _{2} &=&n_{2}\,.
\end{eqnarray}
A solution to these equations always exist, and the extremal 2-brane labeled by $(n_{1},n_{2})\in \mathbb{Z}^{2}$ is parameterized by $a_{0}$ as
\begin{eqnarray}
\eta _{1,\text{extr}} &=&\sinh ^{-1}\left( \frac{a_{0}^{2}-n_{1}^{2}-2n_{1}a_{0}}{a_{0}^{2}}\right) \,,  \notag \\
\eta _{2,\text{extr}} &=&\sinh ^{-1}\left( \frac{-n_{1}n_{2}a_{0}\pm \sqrt{a_{0}^{2}n_{1}^{2}+n_{2}^{2}-1}}{a_{0}^{2}n_{1}^{2}-1}\right) \,.
\end{eqnarray}

Other stable, spinning BPS branes can also be obtained when the bosonic CS matter is included, as discussed in Ref.\cite{EGMZone}.

%%%%%%%%%%%%%%%%%%%%%%%%%%%%%%%%%%%%%%%%%%%% [4]
\section{Codimension-two branes in higher-dimensions}
\label{Codimension2}
%%%%%%%%%%%%%%%%%%%%%%%%%%%%%%%%%%%%%%%%%%%%

Configurations similar to the 2-branes discussed in AdS$_5$ can be generalized to higher-dimensional AdS$_{2p+3}$ spacetime to obtain $2p$-branes. The idea is to introduce suitable identifications in the Euclidean planes, after embedding the AdS$_{2p+3}$ hypersphere, $x\cdot x=-\ell ^{2}$, in the flat space $\mathbb{R}^{2,2p+2}$ with Cartesian coordinates $x^{A}$ ($A=0,\ldots,2p+3$) and signature $(-,+,\cdots ,+,-)$.

A convenient way to obtain a codimension-two brane in higher odd dimensions is to choose the coordinates $x^{0}$, $x^{1}$, $x^{2} $ and $x^{2p+3}$ to represent a three-dimensional $0$-brane (see Section \ref{ThreeD branes}), parameterized by $B\geq 0$, the azimuthal angle $\phi_{12}$, and the time coordinate $\phi_{0,2p+3}\in \mathbb{R}$, thickened by the introduction of $p$ internal radial coordinates, $\tau _{1},\ldots,\tau_{p} \geq 0$, and angles $\phi_{34},\ldots,\phi_{2p+1,2p+2}$, so that \cite{EGMZone}
\begin{equation}
\begin{array}{ll}
x^{0} = A \Upsilon_{1,p} \cos \phi _{0,2p+3}\,,\qquad & x^{1}=B\cos \phi _{12}\,, \\ [0.5em]
x^{2p+3} = A \Upsilon_{1,p} \sin \phi _{0,2p+3}\,, \qquad & x^{2}=B\sin \phi _{12}\,,
\end{array}
\label{012D}
\end{equation}
and for $i = 1,\ldots,p$,
\begin{eqnarray}
x^{2i+1} &=& A \Upsilon_{1,i-1} \sinh\tau_{i} \cos \phi_{2i+1,2i+2}\,,  \notag \\ [0.5em]
x^{2i+2} &=& A \Upsilon_{1,i-1} \sinh\tau_{i} \sin \phi_{2i+1,2i+2}\,.
\label{pairs}
\end{eqnarray}
Here $A^2 = B^{2}+\ell ^{2}$ and we introduced, for compactness, the functions
\begin{equation}
\Upsilon_{l,m} \equiv \prod_{i=l}^m \cosh\tau_{i}\,,
\label{upsilon}
\end{equation}
where $\Upsilon_{1,0} \equiv 1$. We restrict to odd dimensions, but similar transformation exist in even dimensions, as well.

As a result of this construction that preserves the AdS constraint, one obtains a spacetime with locally constant curvature that describes a 2$p$-brane. The metric takes the form
\begin{equation}
ds^{2} = \frac{\ell ^{2}}{A^{2}}\,dB^{2}+B^{2}d\phi_{12}^{2}-A^{2}\,\Upsilon^2_{1,p}\,d\phi _{0,2p+3}^{2} + A^{2} \sum_{i=1}^{p} \Upsilon^2_{1,i-1} \left( d\tau _{i}^{2}+\sinh^{2}\!\tau_{i}\,d\phi_{2i+1,2i+2}^{2}\right) \,.
\end{equation}
In five dimensions ($p=1$), this metric matches the one given by Eq.(\ref{two spins}), with $\tau =\tau_{1}$.

Because some azimuthal angles have deficits with respect to global AdS, conical singularities are generated in the submanifolds that coincide with worldvolumes of static branes, and also causal horizons are generated when the branes are rotating. For example, if only the angle $\phi_{12}$ has a deficit $2\pi( 1-a_0)$, and other angles $\phi_{2i+1,2i+2}$ have period $2\pi$, a static $2p$-brane is produced with conical singularity at the origin, $B=0$; the deficit $a_0$ is related to the negative mass parameter of the brane $M$ \cite{EGMZone}.

As discussed in the previous section, if more angles have deficits, then the brane acquires angular momenta as well, and it might become a system of intersecting $p$-branes. In $2p+3$ dimensions, there can be at most $p+1$ independent/commuting rotations in Euclidean planes of the ($2p+2$)-dimensional spatial section in the embedding space. Therefore, the most general codimension-two brane is described by $p+2$ parameters related to conserved charges, {\it i.e.}, the mass $M$, and angular momenta, $\vec{J}=\left( J_{1},\ldots ,J_{p+1}\right) $. This counting matches the fact that there can be at most $p+2$ angular deficits in the angles $\phi_{0,2p+3}$ and $\phi_{2i+1,2i+2}$, $i=0,\ldots,p$.

The isometry group of AdS$_{2p+3}$, $SO(2,2p+2)$, is of rank $p+2$. This means that the brane can be described by at most $p+2$ independent parameters, each one corresponding to a generator of the Cartan subalgebra. Similarly to the five-dimensional case, the mass parameter is directly related to an angular deficit of a static brane, and $p+1$ angular momenta are related to independent rotations in Euclidean planes.

The angular momenta can be alternatively obtained by applying $p+1$ independent boosts, or Lorentz transformations, to the static brane along independent angles. To be more precise, if $\mathbf{J}_{AB}$ are generators of $SO(2,2p+2)$, then the commuting generators of the Cartan subalgebra are $\mathbf{J}_{0,2p+3}$ and $\mathbf{J}_{2i+1,2i+2}$, where $i=0,\ldots,p$. Geometrically, the angles $\phi_{0,2p+3}$ and $\phi_{2i+1,2i+2}$ are azimuthal angles in the Euclidean planes $x^{0}$-$x^{2p+3}$ and $x^{2i+1}$-$x^{2i+2}$, respectively.

Furthermore, as in the five-dimensional case, the embedding angles $\phi_{0,2p+3}$ and $\phi_{2i+1,2i+2}$ are related to the time coordinate $t$ and $p+1$ periodic angles $\theta_{i}=\left( \phi ,\theta ,\ldots \right) \in \lbrack 0,2\pi )$. The relation between the local  coordinates in AdS space and those in the embedding space is captured by the invertible matrix $U$ that generalizes Eq.(\ref{mix matrix}),
\begin{equation}
\left(
\begin{array}{c}
\phi_{0,2p+3} \\
\phi_{12} \\
\phi_{34} \\
\vdots  \\
\phi_{2p+1,2p+2}
\end{array}
\right) =\left(
\begin{array}{ccccc}
u_{1} & b_{1} & b_{2} & \cdots  & b_{p+1} \\
u_{2} & a_{1} & a_{2} &  & a_{p+1} \\
u_{3} & c_{1} & c_{2} &  & c_{p+1} \\
\vdots  &  &  & \ddots  & \vdots  \\
u_{p+2} & q_{1} & q_{2} & \cdots  & q_{p+1}
\end{array}
\right) \left(
\begin{array}{c}
t/\ell  \\
\theta_{1} \\
\theta_{2} \\
\vdots  \\
\theta_{p+1}
\end{array}
\right) \,.  \label{U}
\end{equation}
In this notation, a static brane is represented by $U_{0}=$ diag$\left( a_{0},a_{0},1,\ldots ,1\right) $, where $\phi _{12}=a_{0}\,\phi $, $\phi _{0,2p+3}=a_{0}t/\ell $ and all other angles are periodic, with periods $2\pi$. Static solutions with different transverse planes also exist, but they simply correspond to different matrices $U_0$ in which the second eigenvalue $a_0$ is moved along the diagonal.

A rotating brane is brought to existence from the static brane by applying at most $p+1$ independent Lorentz transformations $\Lambda$ along the angles $\theta_{i}$, each one introducing a component $J_{i}$. Then, the matrix $U$ in (\ref{U}) can be ``unboosted'' to the static brane as $U=U_{0}\Lambda^{-1}$. Again, $U$ has at most $p+2$ independent components and $\det U=a_{0}^{2}$.

%%%%%%%%%%%%%%%%%%%% [4.1]
\subsection{Sources}
\label{higherD sources}
%%%%%%%%%%%%%%%%%%%%

Angular deficits of the embedding angles $\phi_{ij}$, due to identifications $\theta _{i}\simeq \theta _{i}+2\pi $, are read-off from the matrix elements of $U$ as $a_i,b_i,c_i, \ldots , q_i\in (0,1]$. These entries correspond to the parameters that characterize the Killing vector fields used to identify points in the embedding space,
\begin{equation}
\xi _{i}(x)=2\pi a_{i}\,\mathbf{J}_{12}-2\pi b_{i}\,\mathbf{J}_{0,2n+1}+2\pi c_{i}\,\mathbf{J}_{34}+\cdots +2\pi q_{i}\,\mathbf{J}_{2p+1,2p+2}\,.
\end{equation}
A rotating brane with angular momenta $\vec{J}$ is the quotient of AdS$_{2p+3}$ space by these identifications. The spacetime curvature is constant everywhere. However, as in the previous discussions, the allowed regions of spacetime are those that do not possess closed time-like curves, that is, where all the norms
\begin{equation}
\left\Vert \xi _{i}\right\Vert ^{2} = 2\pi\, ( a_{i}B^{2}-b_{i}A^{2}\,\Upsilon_{1,p}^2 + c_{i}\,A^{2}\sinh ^{2}\tau _{1} + \cdots + q_{i}\,A^{2}\,\Upsilon_{1,p-1}^2 \sinh ^{2}\tau _{p}) \,,
\end{equation}
are non-negative. As in five dimensions, the causal horizon $B=B_*(\tau)$ is a boundary of the ``outer" region, where all $\left\Vert \xi_i\right\Vert^2$ are strictly positive. The interior is removed from the manifold by an additional transformation that identifies $B_*$ as a set of points with infinite curvature, that is, a source of a codimension-two where the brane is.

Explicit expressions for this source can be obtained using the method of nontrivial holonomies, explained in previous sections. In order to analyze non-trivial paths around the horizon, we choose the vielbein as
\begin{eqnarray}
& & e^{0} = A\,\Upsilon_{1,p}\,d\phi _{0,2p+3}\,, \qquad e^{1} = \frac{\ell }{A}\,dB\,, \qquad e^{2} = B\,d\phi _{12}\,, \notag \\ [0.5em]
& & e^{2i+1} = A\,\Upsilon_{1,i-1}\,d\tau _{i}\,, \qquad e^{2i+2} = A\,\Upsilon_{1,i-1}\sinh \tau _{i}\,d\phi_{2i+1,2i+2}\,,  \label{e}
\end{eqnarray}
where $i=1,\ldots ,p$ and $\Upsilon_{l.m}$ has been defined in Eq.(\ref{upsilon}). Then the torsionless spin connection, defined outside the source, has non-vanishing components
\begin{eqnarray}
& & \omega ^{01} = \frac{B}{\ell A}\,e^{0}\,, \qquad \omega ^{12} =-\frac{A}{\ell }\,d\phi _{12}\,,  \qquad \omega ^{1,2i+1} = -\frac{B}{\ell A}\,e^{2i+1}\,, \notag \\ [0.5em]
& & \omega ^{1,2i+2} = -\frac{B}{\ell A}\,e^{2i+2}\,, \qquad \omega ^{0,2i+1} = \sinh \tau _{i}\,\Upsilon_{i+1,p}\,d\phi _{0,2p+3}\,, \nonumber
\end{eqnarray}
and with $i<j$,
\begin{eqnarray}
\omega ^{2i+1,2i+2} &=&-\cosh \tau _{i}\,d\phi _{2i+1,2i+2}\,,  \notag \\ [0.3em]
\omega ^{2i+1,2j+1} &=&\sinh \tau_{j}\,\Upsilon_{i-1,j+1}\,d\tau _{i}\,,  \notag \\ [0.3em]
\omega ^{2i+2,2j+1} &=&\sinh \tau_{i}\,\sinh \tau_{j}\,\Upsilon_{i-1,j+1}\,d\phi _{2i+1,2i+2}\,.  \label{w}
\end{eqnarray}
The AdS connection $\mathbf{A}$ has the form given by Eq.(\ref{AdS_connection}). As discussed above, the integral of $\mathbf{F}$ over an infinitesimal surface that contains the horizon is finite, because the coordinate definitions (\ref{012D}, \ref{pairs}) introduce a curvature singularity: the source of the brane $\mathbf{j}$. This source is located at the causal horizon, that is in general a union of various surfaces $\left\Vert \xi _{i}\right\Vert =0$ coming from different Killing vectors.

The only contribution to the integral $\int_{\Sigma_*} \mathbf{F}$ comes from an Abelian subgroup $\int_{\Sigma_*} d\mathbf{A=}\oint_{\mathcal{C}_*} \mathbf{A}$. In Euclidean planes, the loops around the horizon are parameterized by the azimuthal angles $\phi _{ij}$. When the angles have periods smaller than $2\pi $, then $\oint d\phi _{ij}=2\pi\lambda _{ij}$, where $\lambda _{ij}\in (0,1)$ are the constants whose form depend on the matrix (\ref{U}). When there is no deficit, $\lambda_{ij}=1$, and we can drop this term from the source, as it describes a simple $S^{1}$ rotation, that is an ambiguity in a Killing vector that can always be removed.

Writing these contributions to $\mathbf{F}$ in the $x^{i}$-$x^{j}$ planes as the 2-forms $2\pi \lambda _{ij}\delta \left( \Sigma _{ij}\right) $, we obtain the source of a codimension-two brane in $2n+1$ dimensions,
\begin{eqnarray}
\mathbf{j} &=& \frac{2\pi}{\ell } \left( B_*\,\mathbf{J}_{2,2p+3}-A_*\,\mathbf{J}_{12}\right) \lambda_{12} \delta \left( \Sigma
_{12}\right) \notag \\
&&+\frac{2\pi}{\ell }\left[ \Upsilon_{1,p}\,\left( A_*\,\mathbf{J}_{0,2p+3}+B_*\,\mathbf{J}_{01}\right) + \sum_{i=1}^p\ell \sinh \tau _{i}\,\Upsilon_{i+1,p}\,\mathbf{J}_{0,2i+1}\right] \lambda_{0,2p+3}\,\delta \left( \Sigma _{0,2p+3}\right) \notag \\
&&+\frac{2\pi}{\ell }\,\sum_{i=1}^{p}\left[ \Upsilon_{1,i-1}\sinh \tau_{i}\left( A_*\,\mathbf{J}_{2i+2,2p+3}-B_*\, \mathbf{J}_{1,2i+2}\right) +\ell \cosh \tau_{i}\,\mathbf{J}_{2i+2,2i+1}\right. \notag \\
&&\qquad +\frac{1}{2}\sum_{i,j=1}^{p}\left. \ell \sinh \tau_{i}\,\sinh \tau_{j}\,\Upsilon_{i-1,j+1}\,\mathbf{J}_{2i+2,2j+1} \right] \lambda _{2i+1,2i+2}\,\delta \left( \Sigma _{2i+1,2i+2}\right)\,.
\end{eqnarray}
Not all Dirac delta 2-forms $\delta \left( \Sigma _{ij}\right) $ are independent sources, because $\Sigma _{ij}$ are Euclidean planes in the embedding space $\mathbb{R}^{2,2p+2}$ and $\delta \left( \Sigma _{ij}\right) $ has to be expressed in the AdS coordinates. A more explicit form of $\mathbf{j}$ depends on the number of angular momenta and the identification under consideration.

In the next subsection, we will show that these spinning 2-branes can be stable in the framework of CS supergravity.

%%%%%%%%%%%%%%%%%%%%%%%4.2
\subsection{Killing spinors}
%%%%%%%%%%%%%%%%%%%%%%%

Consider a CS supergravity with $N$ gravitini for some super AdS group. The smallest superalgebras that contain the AdS$_{2p+3}$ are constructed and classified in $2p+3$ dimensions in Ref.\cite{TZ2}. A spinor $\psi_s$ ($s=1,\ldots, N$) with $2^{p+1}$ components, transforms as a vector in some internal subalgebra $so(N)$, $sp(N)$ or $su(N)$, depending on the dimension of spacetime; generically, the superalgebra also contains additional bosonic generators required by its closure.

A $2p$-brane solution of the theory is a BPS configuration if its gauge connection $\mathbf{A}$ corresponds to (\ref{AdS_connection}, \ref{e}--\ref{w}), with all other fields set to zero, and it admits nontrivial Killing spinors $\epsilon_s$ satisfying Eq. (\ref{k-spinor-eq}). As we show below, in order to satisfy this requirement, the brane geometry must have at least one nonzero component of the angular momentum. Killing spinors have been calculated in Ref.\cite{EGMZone}, and they have the form
\begin{equation}
\epsilon _{s}=e^{-\frac{1}{2}\,p(B)\Gamma _{1}}\prod_{i=1}^{p}e^{-\frac{1}{2}\,\tau _{i}\Gamma _{2i+1}}e^{\frac{1}{2}\,\phi _{2i+1,2i+2}\,\Gamma_{2i+1,2i+2}}\,e^{\frac{1}{2}\,\left( \phi _{12}\Gamma _{12}-\phi_{0,2p+3}\Gamma _{0}\right) }\chi_s\,.
\end{equation}
The function $p(B)=\sinh ^{-1}(B/\ell )$ is the same as before, and $\Gamma _{a}$ ($a=0,1,\ldots ,2p+2$) are the ($2p+3)$-dimensional Gamma matrices and $\chi_s$ is a constant spinor. Since the set of matrices $\Gamma _{2i+1,2i+2}=\frac{1}{2}\,\Gamma _{2i+1}\Gamma _{2i+2}$ mutually commute and have eigenvalues $\pm i$, the spinor $\chi_s$ can be chosen as their common eigenvector,
\begin{equation}
\Gamma _{2i+1,2i+2}\,\chi_s=i\chi_s\,,\quad (i=1,\ldots ,p)\,.
\end{equation}
Furthermore, the matrix $\Gamma _{0}$, proportional to the product $\Gamma_{12}\cdots \Gamma _{2p+1,2p+2}$, can be normalized as
\begin{equation}
\Gamma _{0}\chi_s=i\chi_s\,.
\end{equation}
Then, the spinor $\epsilon _{s}$ adopts the form
\begin{equation}
\epsilon _{s}=e^{-\frac{1}{2}\,p(B)\Gamma _{1}}e^{\frac{i}{2}\,\left( \phi_{12}-\phi _{0,2p+3}\right)} \prod_{i=1}^{p}e^{-\frac{1}{2}\,\tau_{i}\Gamma _{2i+1}}e^{\frac{i}{2}\,\phi _{2i+1,2i+2}}\chi_s\,.
\end{equation}

In order to have a globally defined, $\epsilon_s$  it is necessary to impose periodicity conditions in the angles $\theta_i$. From the matrix $U$ with the matrix elements $U_{\mu \nu }^{\lambda }$ in Eq.(\ref{U}), we find that a change $\theta _{j}\rightarrow \theta _{j}+2\pi $ of the $j$-th angle ($j=1,\ldots ,p+1$), induces a change in the embedding angles
\begin{eqnarray}
\phi _{12} &\rightarrow &\phi _{12}+2\pi a_{j}\,,  \notag \\
\phi _{0,2p+3} &\rightarrow &\phi _{0,2p+3}+2\pi b_{j}\,,  \notag \\
\phi _{2i+1,2i+2} &\rightarrow &\phi _{2i+1,2i+2}+2\pi\, U_{2i+1,2i+2}^{j}\,.
\end{eqnarray}
Thus, for fixed $B$, $\tau _{i}$ and $\theta _{i}$ ($i\neq j$), the spinor changes as
\begin{equation}
\epsilon _{s}\left( \theta _{j}+2\pi \right) =e^{i\pi \,\left( a_{j}-b_{j}\right)}
\prod_{i=1}^{p}e^{i\pi U_{2i+1,2i+2}^{j}} \,
\epsilon_{s}\left( \theta _{j}\right) \,.
\end{equation}
Requiring that the spinor be periodic or antiperiodic function of the angles, $\epsilon _{s}\left( \theta _{j}+2\pi \right) =\pm \epsilon_{s}(\theta _{j})$, one obtains $p+1$ independent conditions on the parameters,
\begin{equation}
a_{j}-b_{j}+\sum_{i=1}^{p}U_{2i+1,2i+2}^{j}=N_{j}\,,\quad (N_{j}\in \mathbb{Z\,},\text{ }j=1,\ldots ,p+1)\,.  \label{extremal_D}
\end{equation}
These conditions \emph{linear} in the parameters are always satisfied by the global AdS$_{2p+3}$, for which $U=1$. Also, these conditions are never satisfied by static branes, where there is only one parameter in the interval $(0,1)$. However, with at least two independent parameters, above equations present the extremality conditions between the sets of parameters that are, as we saw before, related to the conserved charges $M$ and $J_{i}$ of the theory. Similarly as in five dimensions, these linear conditions that define the BPS states do not coincide for $p\geq 1$ with the \emph{non-linear} conditions of the extremality of angular momentum.

The extremality conditions (\ref{extremal_D}) applied to five-dimensional 2-branes, reproduce results discussed in Section \ref{BPS in 5D}. The BPS states described here are easily generalized when other forms of CS matter are added, as for example a $U(1)$ field. For static codimension-two branes, this case was analyzed in Ref.\cite{EGMZone}.

%%%%%%%%%%%%%%%%%%%%%%%%%%%% [5]
\section{Conclusions and outlook}
%%%%%%%%%%%%%%%%%%%%%%%%%%%%

This paper is devoted to the study of codimension-two branes in AdS$_{D\geq 3}$ spacetimes. The branes are classified by a set of parameters \{$a_i,b_i, c_i, \ldots \}$ related to their mass and angular momenta. The construction is based on the introduction of Killing vector identifications in the embedding space $\mathbb{R}^{2,D-1}$ of global AdS$_D$, where the above parameters become angular deficits. These identifications have fixed points and produce configurations that are locally AdS, but whose global structure is changed and describe a variety of $2p$-brane configurations.

In this construction, one is faced with entire regions of spacetime that admit closed time-like curves. They are removed by an
additional identification of the causal horizon with a point. The last identification, which is not produced by a Killing vector,
keeps the manifold locally AdS everywhere, except at the point defining the causal horizon, where the Riemann curvature becomes infinite. In the field equations, this is presented as a brane source. The source generically has the form of a Dirac delta
distribution --without derivatives--, contrary to recent reports in the literature (see, {\it e.g.}, \cite{VVone,VVtwo}). This is
possible on account of a non-vanishing torsion at the singular locus defining the position of the brane.

Our construction guarantees that the AdS curvature satisfies $\mathbf{F}=\mathbf{j}$, independently on the form of the action.
This means that, without a source, any action admitting global AdS as a solution can also have these (spinning) branes as exact
solutions. In order to include the sources, a suitable interaction also has to be added in the action.

Geometrically, the manifolds obtained in this article present one or more (maximally $[(D-1)/2]$) spinning codimension-two branes that intersect. Thus, the source is a sum of at most $[(D-1)/2]$ independent Dirac delta distributions, each one associated to one ($D-3$)-brane, or one Killing vector identification in AdS$_D$ space.

This discussion is restricted to the branes that possess a static limit, {\it i.e.}, that reduce to the static ($D-3$)-brane in the
absence of angular momenta. A construction of this class of spinning branes is shown to be equivalent to a boosting along an azimuthal angle of the static brane, in the spirit of Ref.\cite{MTZ}, where a similar analysis has been done for the BTZ black hole
\cite{BTZ,BHTZ}.

Our approach to obtain a higher-dimensional ($D-3$)-brane from the three-dimensional 0-brane is to thicken it (by introducing further internal coordinates), thus it is natural to expect that there exist more than one inequivalent ways to do this kind of embedding. Furthermore, since identifications of AdS space with or without fixed points produce branes or black holes, respectively, it would be quite interesting to study combined identifications where both branes and black holes are formed. All those cases that lead to nontrivial global geometries are left for future discussion. In particular,  a general analysis like the one performed in \cite{BHTZ} seems as necessary as cumbersome.

Extremal spinning branes are not included in this analysis, because they would correspond to a boost whose velocity is equal to the speed of light. Coordinate transformations that exhibit this Killing vector identification are not well-defined in the extremal limit either. The non-extremality condition, however, does not mean that extremal spinning branes do not exist, but that they should be constructed from a different Killing vector identification. Indeed, in three dimensions, extremal spinning branes have been obtained from identifications in AdS$_3$ in Ref.\cite{MZone}. This extremal spinning 0-brane should be thickened and embedded in higher dimensions similarly as in the non-extremal case, in order to give rise to the extremal codimension-two brane in $D$ dimensions.

To understand the geometry of spinning branes that satisfy $\mathbf{F}=0$, one does not need to know an action -- all actions
that have as  particular solution  $\mathbf{F}=0$ will do the job. This could be, for example, any action from the class of Lovelock actions that generalize General Relativity in higher dimensions. In particular, Einstein-Hilbert action with negative cosmological constant has described spinning branes as solutions in any dimension.

Without knowing the action, one cannot calculate conserved charges in the theory, but one can relate the free parameters \{$a_i,b_i, c_i, \ldots \}$ of the configuration to its mass and angular momenta by their comparison to some known asymptotically AdS solution, for example the BTZ black hole. This is possible because the set of independent parameters is equivalent to the set of Casimir invariants, thus equivalent to the gauge invariant charges.

One question that naturally comes to mind in the presence of these $p$-branes which do not affect the local curvature, but modify the topology, is about the experimental tests that could be performed to detect them. A related problem is how would one distinguish these solutions from other configurations by looking at them from far away. It has been recently shown that black holes and naked singularities can be observationally distinguished through their gravitational lensing properties. When used as gravitational lenses, a black hole and a naked singularity of the same mass and symmetries give rise to different  number of images of the same light source. They also possess different orientations, magnification and time delays of images. For more, see, \emph{e.g.}, Refs.\cite{Virbhadra:2002ju,Virbhadra:2007kw}.

In order to analyze the stability of the brane solutions discussed here requires knowing the full theory, as it involves perturbing the fields in the vicinity of the solution $\mathbf{F}=0$, and studying their dynamics from the equations of motion. We have performed this analysis in the case of CS supergravity, where we showed that there are BPS states that preserve a fraction of the original supersymmetry.

Sources can be incorporated in a theory by adding to the action the couplings between the external currents $\mathbf{j}$ and gravitational and matter fields. We are interested in CS AdS supergravity in odd dimensions, where the AdS connection is a
fundamental field. As discussed in \cite{EZ,MZtwo,EGMZone,EGMZtwo}, gauge invariant and background-independent couplings between a $2p$-brane and a connection is provided by the generalization of the minimal coupling, where the role of the vector potential is played by the $(2p+1)$-CS form. The bosonic sector of this supergravity is of the Lovelock type and it has the unique AdS vacuum $\mathbf{F}=0$. This vacuum is a degenerate zero of the CS field equations, that means that linear perturbation analysis is not well-defined in this background. This problem is circumvented in the supersymmetric case, by showing that the solution $\mathbf{F}=0$ preserves some supersymmetries and, using symmetry arguments, that
it must be a stable state whose charges saturate the Bogomolny'i bound in this background.

The situation in which there is no well-defined linear approximation around some very symmetric background is typical for
higher-dimensional CS theories \cite{CTZ}. The problem is originated in the so-called irregular sectors of the phase space, where the canonical Hamiltonian analysis is not applicable. In these sectors, local symmetry increases and the number of physical degrees of freedom decreases \cite{MZirr}. These sectors are not open sets in phase space, they are submanifolds of measure zero, and one can avoid them by adding to the background a suitable bosonic CS matter \cite{MTZone,MTZtwo}. In that way, one can work in the AdS background in CS gravity even without involving supersymmetry.

Recently,  another solution involving Chern-Simons coupling has been discussed in the context of electric-magnetic duality \cite{Bunster:2011gh} . This magnetically charged solution is geometrically like a Dirac string, but it might be also electrically charged.

%%%%%%%%%%%%%%%%%%%%%%%
\section*{Acknowledgments}
%%%%%%%%%%%%%%%%%%%%%%%

The authors would like to thank to G. Giribet for useful discussions.
This work is supported in part by MICINN and FEDER (grant FPA2008-01838), by Xunta de Galicia (Conseller\'{\i}a de Educaci\'on and grant PGIDIT10PXIB206075PR) and by the Spanish Consolider-Ingenio 2010 Programme CPAN (CSD2007-00042).
This work is partially supported by FONDECYT grants 1100755, 1085322, 1100328, 1110102 and by the Southern Theoretical Physics Laboratory� ACT-91 grant from CONICYT. The Centro de Estudios Cient\'ificos (CECs) is funded by the Chilean Government through the Centers of Excellence Base Financing Program of CONICYT.
A.G. was also partially supported by a CONICET-Doctoral Fellowship and O.M. by the PUCV through the project 123.711/2011.
A.G. wishes to acknowledge the University of Santiago de Compostela for their kind hospitality during his visit, funded by a Santander-USC grant.

\appendix{}

%%%%%%%%%%%%%%%%%%%%%%%%%%%%%%%%%%%%%%%%%%%%%%%%%%%%%%%%%%%%%%%%%%%%%%%%%%%%%%%%%%%%%%%
\section{3D curvature from parallel transport of a vector  \label{transport}}
%%%%%%%%%%%%%%%%%%%%%%%%%%%%%%%%%%%%%%%%%%%%%%%%%%%%%%%%%%%%%%%%%%%%%%%%%%%%%%%%%%%%%%%

The Riemann curvature of three-dimensional 0-brane described in Section \ref{spinning 0} can be obtained from parallel transport of a Lorentz vector $V^{a}$ around the point $r=0$. The equation of parallel transport along $B(r)=Const$ reads
\begin{equation}
DV^{c}=\left( \partial _{\phi }V^{c}+\omega _{\phi }^{cd}\,V_{d}\right)
d\phi =0\,,
\end{equation}
or in components
\begin{eqnarray}
0 &=&\partial _{\phi }V^{0}+\frac{bB}{\ell }\,V^{1}\,,  \notag \\
0 &=&\partial _{\phi }V^{1}+\frac{bB}{\ell }\,V^{0}-\frac{a}{\ell }\sqrt{%
B^{2}+\ell ^{2}}\,V^{2}\,, \\
0 &=&\partial _{\phi }V^{2}+\frac{a}{\ell }\sqrt{B^{2}+\ell ^{2}}\,V^{1}\,.
\notag
\end{eqnarray}
General solution of these  equations has the form
\begin{eqnarray}
V^{0}(\phi ) &=&\frac{Bb}{\ell \Omega }\,\left( \beta \cos \Omega \phi
-\alpha \sin \Omega \phi \right) +\frac{a}{bB}\,\sqrt{B^{2}+\ell ^{2}}\gamma
\,,  \notag \\
V^{1}(\phi ) &=&\alpha \cos \Omega \phi +\beta \sin \Omega \phi \,, \\
V^{2}(\phi ) &=&\frac{a}{\ell \Omega }\,\sqrt{B^{2}+\ell ^{2}}\left( \beta
\cos \Omega \phi -\alpha \sin \Omega \phi \right) +\gamma \,,  \notag
\end{eqnarray}
where the angular frequency is
\begin{equation}
\Omega ^{2}(r)=\frac{a^{2}-b^{2}}{\ell ^{2}}\,B^{2}+a^{2}=\frac{r^{2}}{\ell
^{2}}+a^{2}+b^{2}\,.
\end{equation}
If the transport starts in the point $\phi =0$ with the vector $\bar{V}^a$, the integration constants are
\begin{eqnarray}
\alpha &=&\bar{V}^{1}\,,  \notag \\
\beta &=&\frac{1}{\ell \Omega }\left( -bB\bar{V}^{0}+a\sqrt{B^{2}+\ell ^{2}}
\bar{V}^{2}\right) \,, \\
\gamma &=&\frac{bB}{\ell ^{2}\Omega ^{2}}\left( a\sqrt{B^{2}+\ell ^{2}}\bar{V%
}^{0}-bB\bar{V}^{2}\right) \,.  \notag
\end{eqnarray}
After moving along the circle and returning to the initial point, the vector becomes
\begin{equation}
V^{c}(2\pi )=S_{\ d}^{c}\,\bar{V}^{d}\,,
\end{equation}
where the transformation matrix, in the limit $r\rightarrow 0$, reads
\begin{eqnarray}
S =\left(
\begin{array}{ccc}
1+\frac{b^{4}}{a^{4}-b^{4}}\left( 1-\cos \theta \right) & -\frac{b^{2}}{%
\sqrt{a^{4}-b^{4}}}\,\sin \theta & -\frac{a^{2}b^{2}}{a^{4}-b^{4}}\left(
1-\cos \theta \right) \\
-\frac{b^{2}}{\sqrt{a^{4}-b^{4}}}\,\sin \theta & \cos \theta & \frac{a^{2}}{%
\sqrt{a^{4}-b^{4}}}\sin \theta \\
\frac{a^{2}b^{2}}{a^{4}-b^{4}}\left( 1-\cos \theta \right) & -\frac{a^{2}}{%
\sqrt{a^{4}-b^{4}}}\sin \theta & \frac{1}{a^{4}-b^{4}}\left(
-b^{4}+a^{4}\cos \theta \right)%
\end{array}
\right) \,.
\end{eqnarray}
We have introduced the angle
\begin{equation}
\theta =\lim_{r\rightarrow 0}2\pi \Omega =2\pi \sqrt{a^{2}+b^{2}}\,.
\end{equation}

In the static case ($b=0$), we have $\theta _{0}=2\pi a_{0}$ and
\begin{equation}
S_{\text{static}}=\left(
\begin{array}{ccc}
1 & 0 & 0 \\
0 & \cos \theta _{0} & \sin \theta _{0} \\
0 & -\sin \theta _{0} & \cos \theta _{0}
\end{array}
\right) =e^{\theta _{0}\mathbf{J}_{12}}\,.
\end{equation}
Nontrivial curvature is due to the current
\begin{equation}
S_{\text{static}}=e^{-\int \mathbf{j}_{\text{static}}}\,,\qquad \int \mathbf{j}_{\text{static}}=-2\pi a_{0}\,\mathbf{J}_{12}\,.
\end{equation}

In the rotating case ($b\neq 0$), we can rewrite the matrix $S$ as a composition of the 12-rotation and 02-boost,
\begin{equation}
S=e^{\eta \mathbf{J}_{02}}\,e^{\theta \mathbf{J}_{12}}\,e^{-\eta \mathbf{J}%
_{02}}\,,  \label{U as composition}
\end{equation}
with the angles
\begin{eqnarray}
\theta &=&\lim_{r\rightarrow 0}2\pi \Omega =2\pi \sqrt{a^{2}+b^{2}}\,, \\
\tanh \eta &=&\lim_{r\rightarrow 0}\frac{bB}{a\sqrt{B^{2}+\ell ^{2}}}=\frac{%
b^{2}}{a^{2}}\,.
\end{eqnarray}
Again, a spinning brane is obtained from the static one after applying the Lorentz boost,
\begin{equation}
e^{\eta \mathbf{J}_{02}}=\left(
\begin{array}{ccc}
\cosh \eta & 0 & \sinh \eta \\
0 & 1 & 0 \\
\sinh \eta & 0 & \cosh \eta%
\end{array}%
\right) \,.  \label{S,Lambda}
\end{equation}
Nontrivial curvature ($S \neq 1$) is due to the source,
\begin{equation}
S=e^{-\int \mathbf{j}_{\text{curvature}}}\,.
\end{equation}
Because the generators $\mathbf{J}_{12}$ and $\mathbf{J}_{02}$ do not commute, $\left[ \mathbf{J}_{12},\mathbf{J}_{02}\right] =\mathbf{J}_{01}$, we apply the Baker-Campbell-Hausdorff formula,
\begin{equation}
e^{\eta \mathbf{J}_{02}}e^{\theta \mathbf{J}_{12}}e^{-\eta \mathbf{J}_{02}}= e^{ \theta ( \mathbf{J}_{12}\,\cosh \eta -\mathbf{J}_{01}\sinh \eta )} \,,
\end{equation}
and we obtain the source as
\begin{eqnarray}
\int \mathbf{j}_{\text{curvature}} &=&\theta \left( -\mathbf{J}_{12}\cosh \eta +\mathbf{J}_{01}\sinh \eta \right)  \notag \\
&=&\frac{2\pi }{\sqrt{a^{2}-b^{2}}}\left( -a^{2}\mathbf{J}_{12}+b^{2}\mathbf{J}_{01}\right) \,.  \label{j_curv}
\end{eqnarray}

This analysis is incomplete because the full source contains a torsional part as well, $\mathbf{j}=\mathbf{j}_{\text{curvature}} +\mathbf{j}_{\text{torsion}}$.

%%%%%%%%%%%%%%%%%%%%%%%%%%%%%%%%%%%%%%%%%%%%%%%%%%%%%%%%%%%%%%%%%%%%%%%%%%%%%%%%%%%

\end{document}